\documentclass[amsmath,amssymb,twocolumn,aps,prb]{revtex4-2}
\usepackage{graphicx}
\usepackage{epstopdf}
\usepackage{bm}% bold math
\usepackage{subfigure}
\usepackage{sidecap}

\usepackage{graphicx}
\usepackage{color}
\usepackage{epstopdf}
\usepackage{amssymb}
\usepackage{amsmath}

\usepackage{bm}% bold math
\graphicspath{{Figures/}}
\usepackage{subfigure}
\usepackage{sidecap}
\usepackage {times}

\newcommand{\al}{\alpha}

\newcommand{\ga}{\gamma}

\newcommand{\lam}{\lambda}

\newcommand{\De}{\Delta}

\newcommand{\be}{\begin{equation}}
\newcommand{\ee}{\end{equation}}

\newcommand{\bea}{\begin{eqnarray}}
\newcommand{\eea}{\end{eqnarray}}
\newcommand{\bd}{\begin{displaymath}}
\newcommand{\ed}{\end{displaymath}}
\newcommand{\ba}{\begin{array}}
\newcommand{\ea}{\end{array}}
\newcommand{\bi}{\begin{itemize}}
\newcommand{\ei}{\end{itemize}}
\newcommand{\bc}{\begin{center}}
\newcommand{\ec}{\end{center}}
\newcommand{\bfl}{\begin{flushleft}}
\newcommand{\efl}{\end{flushleft}}
\newcommand{\bfr}{\begin{flushright}}
\newcommand{\efr}{\end{flushright}}
\newcommand{\non}{\nonumber}

\newcommand{\bl}{\begin{aligned}}
\newcommand{\el}{\end{aligned}}

\newcommand{\hP}{\hat{P}}

\newcommand{\hh}{\hat{h}}

\newcommand{\hDe}{\hat{\Delta}}

\newcommand{\hchi}{\hat{\chi}}

\newcommand{\bJ}{\bar{J}}

\newcommand{\blam}{\bar{\lambda}}

\newcommand{\tiP}{\tilde{P}}

\newcommand{\fs}{\frac{1}{2}}

\newcommand{\ra}{\rangle}
\newcommand{\la}{\langle}

  \def\bq{{\bf q}}

\def\bQ{{\bf Q}}   
  
 \def\bd{{\bf d}} \def\bS{{\bf S}} \def\bJ{{\bf J}}
 \def\bS{{\bf S}}

 \def\bJ{{\bf J}}

\def\={\!\!\!&=&\!\!\!}
\def\+{\!\!\!&&\!\!\!+~}
\def\-{\!\!\!&&\!\!\!-~}

\usepackage{color}
\usepackage[dvipsnames]{xcolor}
\usepackage{xcolor}
\def\BLU{\color{black}}  
\def\BLN{\color{black}}  
\usepackage[colorlinks=true,citecolor=blue]{hyperref}

\usepackage{hyperref,cleveref}
\begin{document}

\title{
%Nuclear specific heat in singlet ground state magnets
Hyperfine coupling in singlet ground state magnets}

\author{Peter Thalmeier}
\affiliation{Max Planck Institute for Chemical Physics of Solids, D-01187 Dresden, Germany}
\date{\today}

\begin{abstract}
The influence of hyperfine coupling to nuclear spins and of their quadrupolar splitting on
the induced moment order in singlet ground state magnets is investigated. The latter
are found among non-Kramers f electron compounds.
Without coupling to the nuclear spins these magnets have a quantum critical point (QCP)  separating
paramagnetic and induced moment regime. The hyperfine interaction suppresses the QCP and
leads to a gradual crossover between induced electronic and nuclear hyperfine coupling dominated magnetic order.
It is shown how the critical temperature depends on the electronic and nuclear control parameters
including the nuclear spin size and its possible nuclear quadrupole splitting. In particular the dependence
of the specific heat on the control parameters and applied field is investigated for ferro- and antiferromagnetic order. It is shown that the three peak structure in the electronic induced moment regime gradually changes to a two-peak structure in the hyperfine coupling dominated nuclear moment order regime or for increasing field strength. Most importantly the possibility of a reentrance behavior of magnetic order or likewise nonmonotonic critical fields due to hyperfine coupling influence is demonstrated.
Finally the systematic evolution of the phase diagram under the influence of nuclear quadrupole coupling is clarified.
\end{abstract}
%\pacs{ }
\maketitle

\section{Introduction}
\label{sec:intro}

The common picture of semiclassical magnetism involves ions with a  degenerate spin or pseudospin ground state
which carry free magnetic moments in the paramagnetic state.  They then order spontaneously due to intersite interactions
at the magnetic transition temperature \cite{majlis:07}. The size of the ordered saturation moment will be reduced from the classical value by the effect of quantum fluctuations and possibly by geometric or interaction frustration  \cite{schmidt:17}. This simple picture is applicable to numerous compounds with transition elements, in particular the 4f-based materials. For these the presence of the crystalline electric field (CEF) plays an essential role. For half integer total angular momentum J of the 4f ions the  ground state has always Kramers degeneracy implying picture of  semiclassical order  may be used, except in the quasi-one dimensional case. But for integer J non-Kramers ions such as Pr, Tm, Tb a distinct possibility arises because the CEF ground state may be a singlet $|0\ra$  that is {\it nonmagnetic}, i.e.$\la 0|\bJ|0\ra=0$, preventing the application of the quasiclassical theory of magnetic order. However, if  there is a close by excited state e.g. another nonmagnetic singlet $|1\ra$  the two states can be connected by a finite nondiagonal matrix element $m_s=\la 0|J_z|1\ra$ of a total angular momentum component. Then the intersite exchange of CEF states may spontaneously mix the two states and thus induce
a moment in the superposition ground state of $|0\ra, |1\ra$  that forms below the induced moment ordering temperature. In this mechanism the creation and ordering of moments happens simultaneously at the transition temperature. Literally this picture of induced moment order applies if the other CEF states happen to lie at much higher energy.

{\BLU The theory of induced magnetic ordering for singlet ground state state systems has been systematically presented
in previous publications where more general cases like three-singlet models \cite{thalmeier:21} or singlet doublet-models on the honeycomb lattice with nontrivial topology were investigated \cite{akbari:23}. Furthermore a comparative study of models with singlet, doublet and triplet excited states has been performed, in particular the conditions for  existence of critical 'soft-mode' magnetic excitons and their connection to induced order were deduced \cite{thalmeier:24}. Finally the theory of magneto- and barocaloric properties of the singlet-singlet induced magnets and field dependence of critical magnetic excitons was developed in \cite{thalmeier:25}. We note that it is also possible that higher hidden order (HO) multipole moments  like quadrupoles \cite{takimoto:08} or hexadecapoles \cite{haule:10} may appear through the induced moment mechanism.
However, these previous treatments involved only f- electronic (CEF state) degrees of freedom ('e-dominated region')  and neglected the influence of coupling to nuclear spin states which may become important at very low temperatures compared to the CEF splitting, in particular for control parameters around the QCP (cooperative 'e/n region').} Indeed close to the QCP where the induced ground state moment collapses with a square root singularity \cite{thalmeier:24} the ordering is  naturally very sensitive to external perturbations like magnetic field \cite{thalmeier:25} or pressure \cite{jensen:91}.
A similar sensitivity is caused by further internal interactions like the hyperfine coupling to nuclear spins and the splitting of the latter originating from a nuclear quadrupole potential. {\BLU Further below the QCP ('n-dominated region') the hyperfine coupling becomes the driving force for magnetic ordering. The present work supplements the previous purely electronic CEF singlet-singlet model with the hyperfine and quadrupolar terms of nuclear spins and investigates their influence on the systematic variation of magnetic transition temperature, specific heat, magnetocaloric properties and phase-boundaries with a peculiar type of reentrant behavior.}
Mostly these nuclear hyperfine effects have been investigated for
4f magnetic compounds with Kramers ground state ions like Ce or Yb \cite{steppke:10,steinke:13,knapp:23,knapp:25,gabani:25} or non-Kramers ions with degenerate ground states like Pr \cite{aoki:11} or Ho \cite{kitazawa:25} cage compounds. By applying magnetic fields the ordered phases of such compounds may be tuned to quantum critical points which are also influenced by the presence of hyperfine coupling effects\cite{eisenlohr:21}.

 The hyperfine interaction may indeed be strong for some of the non-Kramers ions which have possible CEF singlet ground states, depending on local site symmetry and CEF potential of the 4f compound in which they appear. These ions and some of their essentials, including the strength of hyperfine couplings are listed in Table~\ref{tbl:REdata}.
Their importance for compounds where non-Kramers rare-earth (RE) ions  have a nonmagnetic singlet ground state has  in fact been known for some time \cite{andres:73,murao:79,kubota:80,murao:81,moller:82,jensen:91,miki:92} and has gained renewed attention more recently \cite{zajarniuk:22,bangma:23,zic:25}. In  particular concerning the magnetocaloric properties and anomalous critical field dependence on temperature which may be interpreted as reentrance of multipolar or magnetic order. The latter effect has also been seen in  Kramers ion Yb compounds \cite{knapp:25} for magnetic order and was attributed to the effect of  hyperfine coupling. \\

In this work we investigate in detail how the hyperfine coupling and quadrupolar splitting of nuclear spins can be included in the induced moment mechanism for singlet ground state magnets. We reconsider the phase diagram as function of electronic (CEF) and nuclear control parameters and discuss how the QCP is suppressed by the nuclear interactions and changed into a gradual crossover  from CEF exchange dominated induced moment magnetism to hyperfine coupling dominated nuclear magnetism.
Particular attention is given to the specific heat anomalies (peak signatures)  that have both electronic CEF and nuclear contributions. It is demonstrated that depending on the location of control parameters a triple peak or double peak structure of the specific heat is expected as function of temperature. Furthermore we discuss the magnetic field dependence of specific heat of singlet ground state ferro- (FM) and antiferromagnetic (AFM) cases with nuclear spins involved and identify their characteristic signatures on both sides of the QCP. Most importantly we show that for the AFM case the coupled system of split electronic singlets and nuclear spins leads to a reentrance of magnetic order as function of temperature for high fields  This observation and the equivalent anomalous nonmonotonic critical field dependence on temperature is a direct consequence of the hyperfine coupling. {\BLU Finally we show that the crossover between electronic and nuclear ordering regimes is quite sensitive to the sign and size of the nuclear quadrupole interaction.}

\section{Hyperfine coupling for singlet ground state induced moment systems}
\label{sec:hyperfine}

For investigating the effect of nuclear hyperfine coupling on induced electronic
magnetic order and the resulting signatures in the specific heat we use the
most simple and yet  illustrative and experimentally relevant singlet-singlet model coupled to nuclear spins (nSSM), consisting of transverse Ising type electronic part and concomitantly an Ising type nuclear hyperfine coupling term.
It is a nuclear-spin extension of the induced moment model investigated
previously \cite{thalmeier:24,thalmeier:25} for various thermodynamic and dynamic properties.
%
% %%%%%%%%%%%%%%%%%%%%% figure %%%%%%%%%%%%%%%%%%%%%%%%%%%%
\begin{figure}
%\vspace{1cm}
\includegraphics[width=1.1\columnwidth]{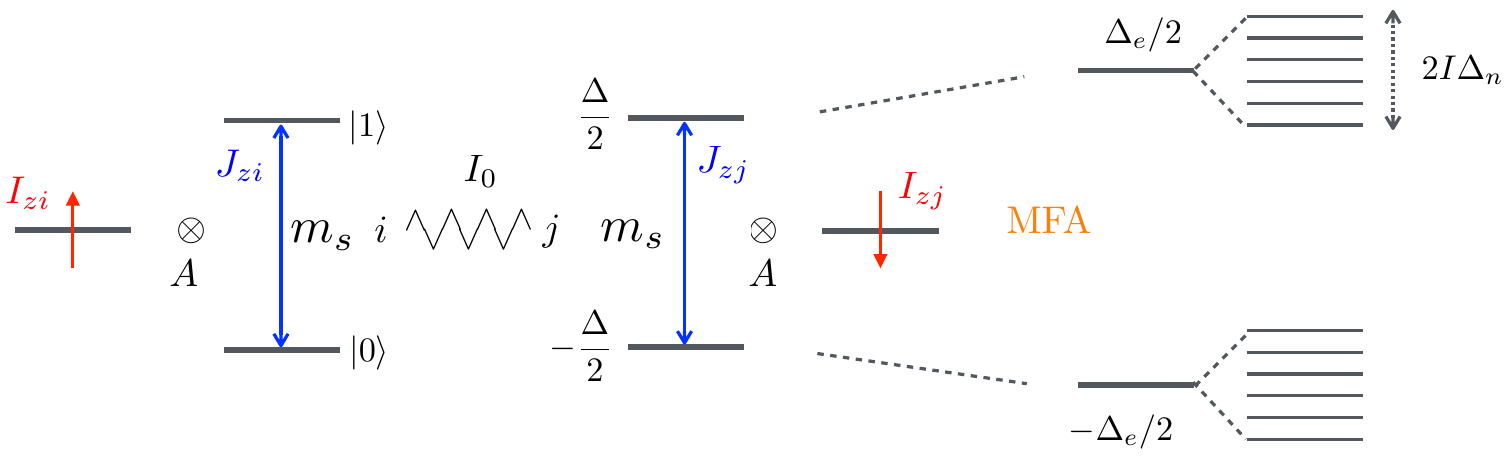}
\caption{{\BLU Schematic presentation of the singlet-singlet electronic $(J_z=m_sS_x)$ and nuclear spin $(I_z)$
 interaction model. Their on-site hyperfine interaction $A$ and intersite electronic exchange $I_0$ leads to
 (FM or AFM) order causing the  shifting and splitting of molecular field levels on the right side.}}
\label{fig:fig1}
\end{figure}
%%%%%%%%%%%%%%%%%%%%%%fig%%%%%%%%%%%%%%%%%%%%%%%%%%%%%%%
%
%
% %%%%%%%%%%%%%%%%%%%%% figure %%%%%%%%%%%%%%%%%%%%%%%%%%%%
\begin{figure}
%\vspace{1cm}
\includegraphics[width=0.99\columnwidth]{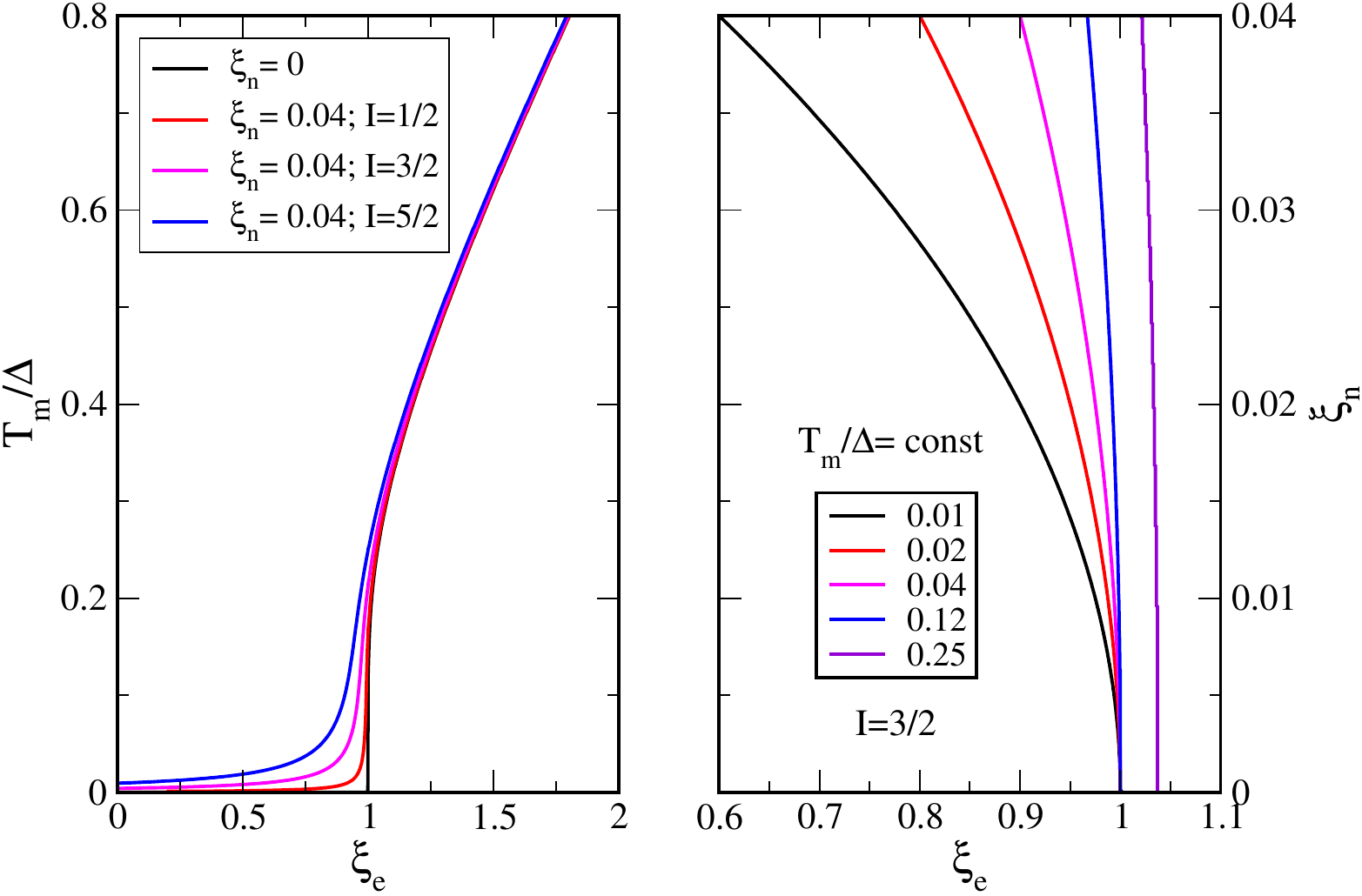}
\caption{Left panel: Critical temperature T$_m$ for induced magnetic order (FM or AFM) as function of electronic $\xi_e$ for fixed nuclear $\xi_n$ and three different sizes of nuclear spin I. The black line is for vanishing $\xi_n$ when induced order appears only above the QCP at $\xi_e=1$ (Eq.~(\ref{eq:Tm0})). For finite $\xi_n$ T$_m$ is finite for any $\xi_e$, {\BLU progressively suppressing the QCP with increasing $\text{I}$.
Right panel: Complementary contour plot of $T_m/\De= \text{const}$ in $\xi_e$-$\xi_n$ plane.}}
\label{fig:fig2}
\end{figure}
%%%%%%%%%%%%%%%%%%%%%%fig%%%%%%%%%%%%%%%%%%%%%%%%%%%%%%%
%
The electronic states consist of two slightly split CEF singlets $|0\ra$ and $|1\ra$ at energies $\pm\De/2$, respectively, where $\De$ is the splitting energy.  They  may originate as low lying states for non-Kramers ions with total angular momentum $J=4,6$ which are split off from the group of  higher lying states by a suitable crystalline electric field. As an example we can have such non-Kramers quasi-doublet in a uniaxial CEF (e.g. $D_{4h}$ symmetry)  for total angular momentum J=4. Then the SSM  corresponding to a $D_{4h}$ representation may be given in terms of free ion states $|J,M\ra\equiv|M\ra$ $(|M|\leq 4)$  by
\be
\bl
%&%\mbox{SSM:} \\
&|0\ra = \cos\al |0\ra  +\sin\al\frac{1}{\sqrt{2}} (|+4\ra+|-4\ra) \;\;(\Gamma^{(1)}_1)
\\
&|1\ra =\frac{1}{\sqrt{2}}(|+4\ra-|-4\ra) \;\; (\Gamma_2)
\quad\quad\quad
\quad\quad\quad
\quad
\el
\label{eq:SSM}
\ee
This Ising-type SSM is relevant for Pr compounds \cite{jensen:91,thalmeier:24} with uniaxial symmetry 
 and has in also been applied to U- compounds~\cite{santini:94,thalmeier:02,sundermann:16}.
 The mixing angle $\al$ is determined by the CEF parameters.
% In this case the CEF mixing angle $\theta$ is close to $\pi/2$.
 The Ising moment operator $J_z=m_sS_x$ in pseudo spin representation within this subspace where
$ m_s=\la 0|J_z|1\ra =8\sin\al$ is then given by
\be
\bl
&
J_z
\!=\!
\frac{m_s}{2}
\left(
 \begin{array}{cc}
0& 1\\
1& 0
\end{array}
\right)=m_sS_x;
\label{eq:Jssm}
\el
\ee
where $\bJ$- operators refer to the free $|JM\ra$ states and $\bS$ are the pseudo-spin operators in the reduced subspace of  CEF singlet states. We note that a singlet-singlet level scheme in a uniaxial symmetry is not able  support an xy-type SSM.
An inspection of the point group multiplication tables leads to the conclusion that this is forbidden for any symmetry \cite{thalmeier:24}.\\

The nuclear spin states  $|I I_z\ra\equiv |I_z\ra$ of the magnetic ion with spin size I form a (2I+1)- fold degenerate multiplet.
It will be split in the magnetically ordered phase due to the action of molecular field mediated by the hyperfine coupling
with strength A in Table~\ref{tbl:REdata}, here assumed to be positive.
Later we also include possible single site nuclear quadrupole terms which will also split the nuclear multiplet. We do not consider the effect of long range nuclear dipole interactions.\\

The nuclear extended nSSM in  pseudospin representation is then defined by the Hamiltonian 
\bea
H&=&\De\sum_iS_{zi}-m_s^2I_0\fs\sum_{\la ij\ra}S_{xi}S_{x_j}\non\\
&+&A\sum_iS_{xi}I_{zi}-m_sh\sum_iS_{xi}
\label{eq:Ham}
\eea
where the first term is the CEF splitting, the second one the effective nearest neighbor (N.N.)  intersite exchange  of 4f- pseudospins,
the third one is the hyperfine coupling which is also necessarily of Ising type since $S_{x,y}$ components are absent in the two-singlet space of Eq.~(\ref{eq:SSM}). The last term is the Zeeman energy of electronic states in external field H with $h=g_J\mu_BH$. The direct Zeeman term of nuclear spins is not included due to the much smaller nuclear magneton $\mu_n\ll\mu_B$. 
{\BLU The pseudospins and nuclear spins are for simplicity placed on a 3D simple cubic lattice, then the Fourier transform of  the inter-site exchange
coupling (second term) is given by $I_\bq=zI_0\gamma_\bq$ where $z=6$ is the coordination number and $\gamma_\bq=(2/z)\sum_i\cos q_i$ $(i=x,y,z)$ the structure factor (lattice constant $a\equiv 1$). This type of n.n. exchange leads either to FM $(\bq=(0,0,0), I_0>0)$  or AFM  $(\bq=(\pi,\pi,\pi), I_0<0)$ order in which cases only the effective exchange $I_e=z|I_0|$  occurs in the thermodynamic and static properties.} Note that we use the convention $I_0$ being positive or negative for FM or AFM exchange, respectively, while $A>0$ means the electronic and nuclear spins are oriented oppositely.\\

In molecular field approximation (MFA) we then have $H_{mf}=\sum_{i\lam}h_{mf}^\lam(i)$ with
\bea
h^\lam_{mf}(i)&=&\hh^\lam_{mf}(i)+\fs\sigma m_s^2 I_e \la S_x\ra_\lam \la S_x\ra_{\blam}
-Am_s\la S_x\ra_\lam\la I_z\ra_\lam\non\\
\hh^\lam_{mf}(i)&=&\De S_z(i)-m_sh_e^\lam S_x(i)+h_n^\lam I_z(i),
\label{eq:MFHam1}
\eea
Here $\lam=A,B$ ($\blam=B,A$)  denotes the sublattices in the case of AFM order with $ \la S_x\ra_{\blam}=-\la S_x\ra_{\lam}$ and likewise   $ \la I_z\ra_{\blam}=-\la I_z\ra_{\lam}$  for zero field. The sublattice index is to be suppressed for FM order. The electronic and nuclear molecular fields (MF) are given by
\bea
h_e^\lam&=&h+\sigma m_sI_e\la S_x\ra_{\blam}-A\la I_z\ra_\lam\non\\
h_n^\lam&=&m_sA\la S_x\ra_\lam
\label{eq:molfield}
\eea
where $\sigma=\pm 1$ for FM and AFM, respectively. We note that the nuclear Zeeman term in Eq.~(\ref{eq:Ham}) which would give a direct contribution $g_n\mu_nH$ to $h^\lam_n$ above is neglected \cite{frossati:76}, assuming $g_n\mu_nH\ll m_sA\la S_x\ra_\lam$ in the present context.
Because the hyperfine term does not mix nuclear spin states in this Ising type model the Hamiltonian will factorise
for each $\tau=I_z$ subspace $(|\tau|\leq I)$ and then we obtain 
the coupled electronic-nuclear MF energy levels given by
\bea
E_{\lam s}^\tau=\frac{\De}{2}\bigl[s(1+\gamma^{'2}_{e\lam})^\fs+2\tau\gamma'_{n\lam}\bigr]
\label{eq:levelenergy0}
\eea
where we defined the dimensionless MF's are defined as $\ga'_{e\lam}=m_sh_e^\lam/\De\equiv :\hDe_{e\lam}$ 
and  $\ga'_{n\lam}=h_n^\lam/\De \equiv :\hDe_{n\lam}$. Furthermore $s=\pm1$ refers to the two MF shifted singlet states originating from original $|1\ra, |0\ra$ states, respectively.
The second term describes the splitting of nuclear spin states by the MF.
The eigenstates are still product states of electronic and nuclear states and therefore
\bea
|E_{\lam +}^\tau\ra&=&cos(\theta_\lam)|1\tau\ra-sin(\theta_\lam)|0\tau\ra\non\\
|E_{\lam -}^\tau\ra&=&sin(\theta_\lam)|1\tau\ra+cos(\theta_\lam)|0\tau\ra
\eea
with the mixing angle given by $\tan 2\theta_\lam=\gamma'_{e\lam}$. The spectrum illustrated in Fig.~\ref{fig:fig1}  is highly symmetric which facilitates further analytical treatment. In the basis of MF eigenstates the two-singlet pseudo spin is represented
by 
\bea
S'_x
&&=
\frac{1}{2}
(1+\ga^{'2}_{e\lam})^{-\fs}
\left(
 \begin{array}{cc}
 -\ga'_{e\lam}& 1\\
1& \ga'_{e\lam}
\end{array}
\right)
\label{eq:Strans}
\eea
Since the nuclear spin states are not mixed by the Ising-like hyperfine term nuclear spin 
operator $I_z$ is unchanged.

% %%%%%%%%%%%%%%%%%%%%% figure %%%%%%%%%%%%%%%%%%%%%%%%%%%%
\begin{figure}
%\vspace{1cm}
\includegraphics[width=0.95\columnwidth]{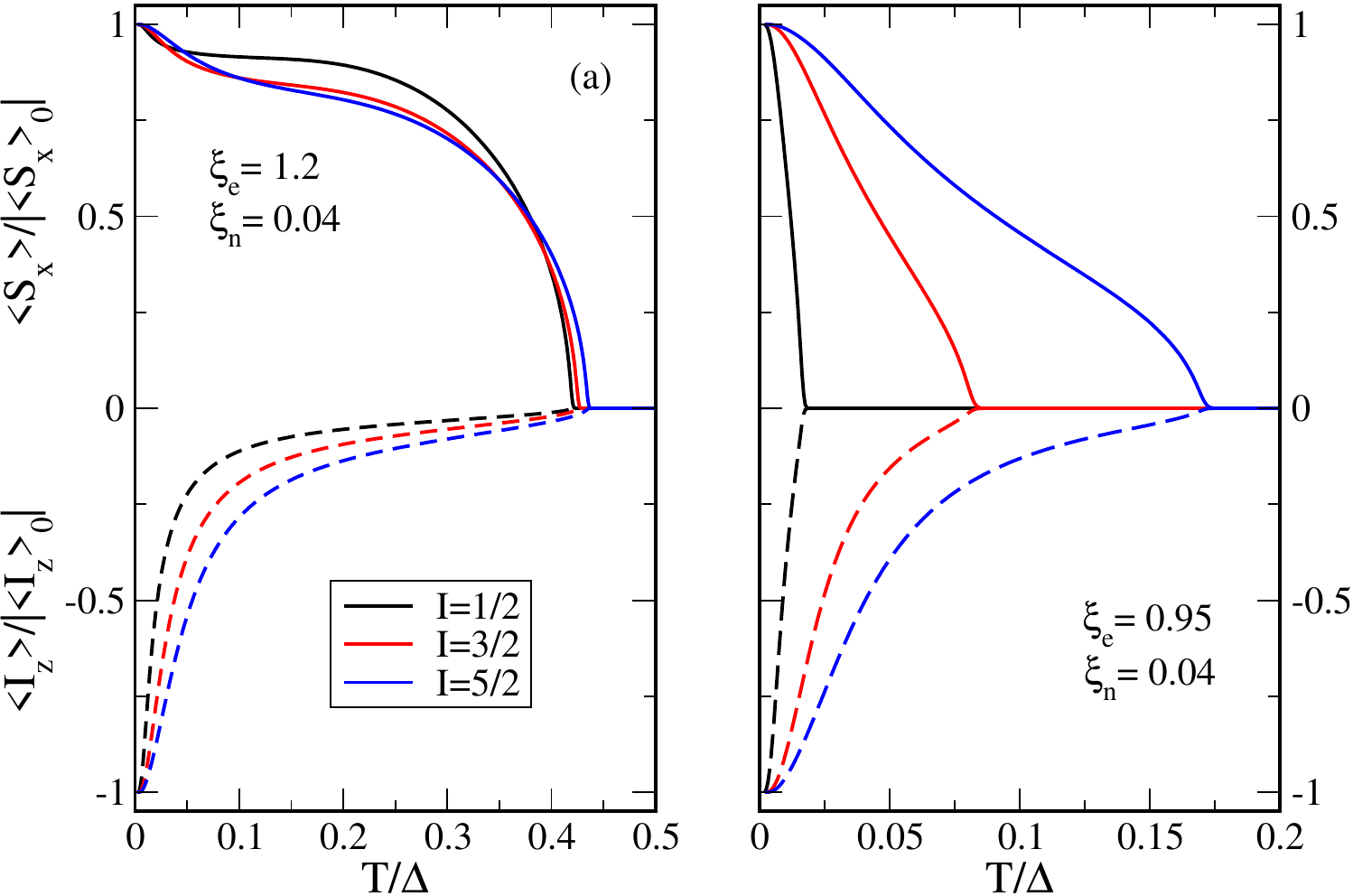}
\includegraphics[width=0.95\columnwidth]{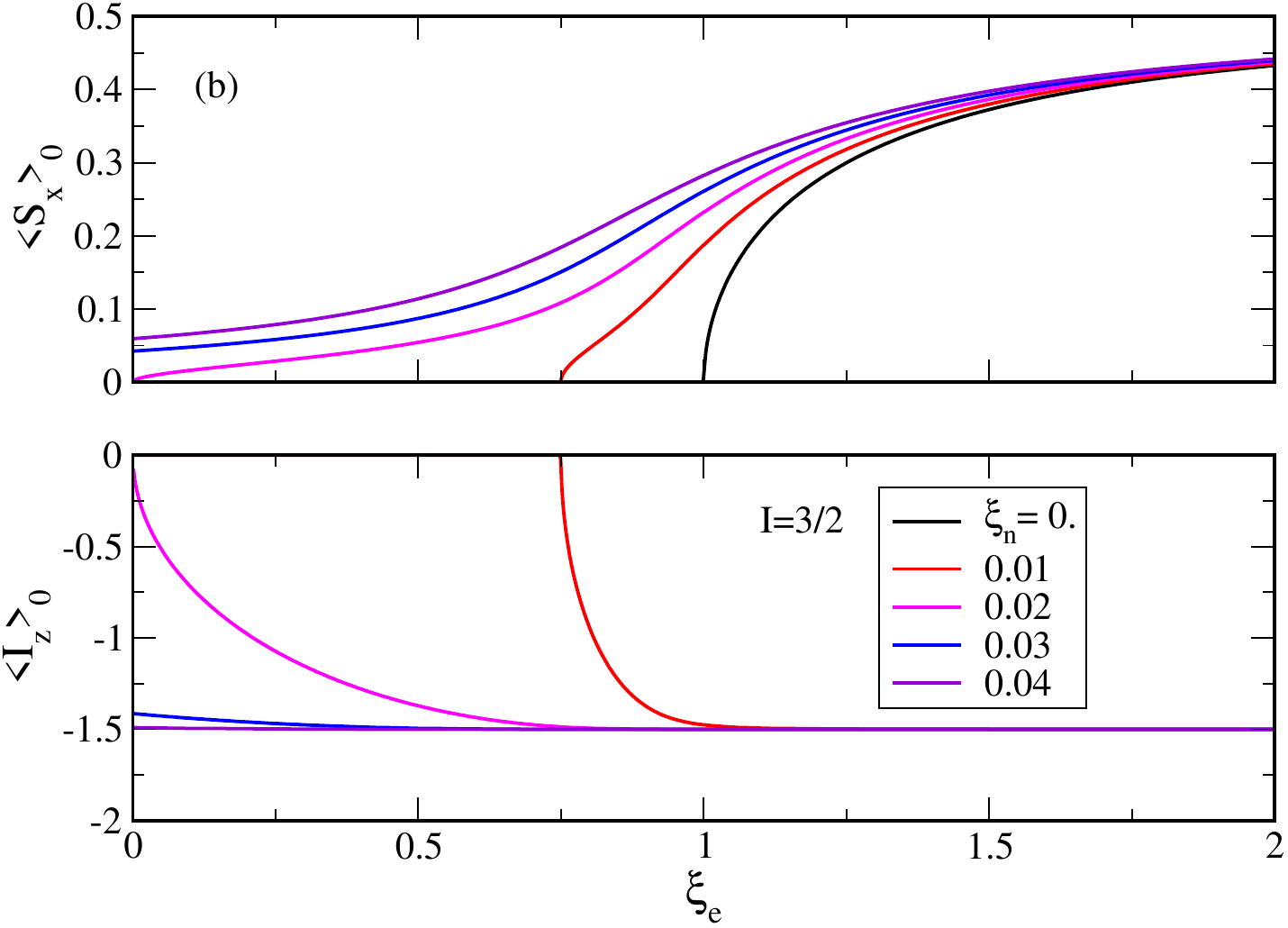}
\caption{(a) Dependence of (normalized) electronic and nuclear order parameters on temperature for hyperfine coupling 
$\xi_n=0.04$. Left panel: {\BLU For  $\xi_e$ above the QCP $\xi_e^c=1$ and various I. Differences in $T_m$ for the three cases correspond to Fig.~\ref{fig:fig2} above the QCP. Right panel: For $\xi_e$ below the QCP when order is due to finite $\xi_n$. Here $\la S_x\ra$,  $\la I_z\ra$ exhibit similar T- dependence  in contrast to (a) but $T_m$ now depends strongly on I.} (b) Electronic and nuclear saturation moments $(T/\De=0.001)$ as function of control parameters. When $T<T_m(\xi_e,\xi_n)$ (Fig.~\ref{fig:fig2}) $\la S_x\ra_0$ is finite and $\la I_z\ra\simeq  -I$. Otherwise both drop to zero.}
\label{fig:fig3}
\end{figure}
%%%%%%%%%%%%%%%%%%%%%%fig%%%%%%%%%%%%%%%%%%%%%%%%%%%%%%%

\section{The critical temperature for induced order of electronic and nuclear moments}
\label{eq:crittemp}

Before solving the MF selfconsistency equations for the order parameters we can determine the phase boundary,
i.e. the critical temperature $T_m(\xi_e,\xi_n,I)$ as function of control parameters $\xi_e,\xi_n$ introduced below and nuclear spin size $\text{I}$ by starting from the paramagnetic regime. It is obtained from the divergence of the static RPA susceptibility
at the critical temperature for the zone center or zone boundary wave vectors of the FM or AFM structure, respectively. 
The static single ion susceptibility for a multilevel system corresponding to operator 
$O=J_z (=m_sS_x),I_z$ is generally given by
\bea
\chi^O_0(T,h')&=&\sum_{E_\mu\neq E_{\mu'}}|\la \mu|O|\mu'\ra |^2\frac{p_\mu-p_{\mu'}}{E_{\mu'}-E_\mu}\non\\
&+&\frac{1}{T}\bigl[\sum_{E_\mu=E_{\mu'}} |\la \mu|O|\mu'\ra |^2p_\mu-\la O\ra^2\bigr],
\label{eq:statsus0}
\eea
where $|\mu\ra=|s\ra\otimes|\tau\ra$ are the paramagnetic states with $s=0,1$ denoting the singlets and $\tau= -I..I$ the nuclear  spin states and $E_\mu,p_\mu$ their energies and thermal populations (App.\ref{sec:occfunc}), respectively. There is no mixing or splitting of these states in the paramagnetic regime; then simply the susceptibilities
are decoupled with
\bea
\chi_0^J=\frac{m_s^2}{2\De}\tanh\frac{\De}{2T}; \;\;\;\;\chi_0^I=\frac{\frac{1}{3}I(I+1)}{T}
\label{eq:parsus0}
\eea
The \bq- dependent  RPA susceptibility corresponding to the Hamiltonian in Eq.~(\ref{eq:Ham})
is then obtained for both operators as
\bea
\chi_\bq^{J,I}=\frac{\chi_0^{J,I}}{1-\chi_0^J(I_\bq+A^2\chi_0^I)}
\label{eq:susRPA}
\eea
\vspace{0.2cm}
with $I_\bq$ defined below Eq.~(\ref{eq:Ham}).
The susceptibility now includes the effect of the hyperfine coupling proportional to A. 
The transition temperature to the induced magnetic order is then given by the condition of vanishing denominator in Eq.~(\ref{eq:susRPA}). This leads to the implicit equation for $T_m(\xi_e,\xi_n,I)$:
\bea
\bigl[\xi_e+\frac{1}{3}I(I+1)(2\xi_n)^2\frac{\De}{2T_m}\bigl]\tanh(\frac{\De}{2T_m})=1
\label{eq:Tm}
\eea
Here we introduced the dimensionless electronic and nuclear control parameters
\bea
\xi_e=\frac{m_s^2I_e}{2\De};\;\;\;\xi_n=\frac{m_s A}{2\De}
\label{eq:control}
\eea
which characterise intersite exchange and hyperfine coupling strength, respectively.
For vanishing hyperfine coupling $(\xi_n=0)$ one obtains explicitly the purely electronic
induced moment transition temperature
\bea
T_m^0(\xi_e)=\frac{\De}{2\tanh^{-1}\bigl(\frac{1}{\xi_e}\bigr)}
\label{eq:Tm0}
\eea
This is shown in Fig.~\ref{fig:fig2} (black line)  together with the general numerical solution of Eq.(\ref{eq:Tm}) for nonzero hyperfine coupling and it will be further discussed in Sec.~\ref{sec:discussion}. Importantly the QCP at $\xi_e=1$ for vanishing hyperfine control parameter $\xi_n$ is now replaced by a crossover for finite $\xi_n$ from a high $T_m$ CEF induced moment region where $T_m\approx T_m^0$ to a low $T_m\ll\De$ nuclear moment dominated ordering region (see also Fig.~\ref{fig:fig2}). For $\xi_e\rightarrow 0$  we have  the asymptotic form derived from Eq.~(\ref{eq:Tm}) as
$T_m \simeq(2/3)\De I(I+1)\xi_n^2(1-\xi_e)^{-1}$.\\

%
%\bea
%T_m \simeq\frac{2}{3}\De I(I+1)\xi_n^2(1-\xi_e)^{-1}
%\label{eq:tmlimit}
%\eea
%
%which approaches a constant independent of $\xi_e$ in this limit.
\label{sec:MFA}

From the derivations in Sec.~\ref{sec:hyperfine} we can now obtain the selfconsistent MF
equations for electronic and nuclear moments as
\bea
\la S_x\ra_\lam&=&\fs\frac{1}{\hDe_e}\bigl[m_sh'+2(\sigma\xi_e\la S_x\ra_{\blam}
-\xi_n\la I_z\ra_\lam)\bigr]\hP_{e\lam}(T)\non\\
\la I_z\ra_\lam&=&-\fs\hP_{n\lam}(T;\la S_x\ra_\lam)
\label{eq:MFA}
\eea
with the thermal occupation functions $\hP_{(e,n)\lam}$ defined by (Eq.~(\ref{eq:popfunc})) in App.~\ref{sec:occfunc}.
Using Eq.~(\ref{eq:molfield}) the corresponding  molecular fields, normalised to $\De$, may  be expressed as
\bea
\gamma'_{e\lam}&=&m_sh'+2(\sigma\xi_e\la S_x\ra_{\blam}-\xi_n\la I_z\ra_\lam)\non\\
\gamma'_{n\lam}&=&2\xi_n\la S_x\ra_\lam
\label{eq:molfield1}
\eea
where  $h'=h/\De$ is the dimensionless field strength.
With these expressions  the MF singlet splitting and nuclear spin splittings are given by
\bea
\hDe_{e\lam}=(1+\ga^{'2}_{e\lam})^\fs;\;\;\;  \hDe_{n\lam}=2\xi_n\la S_x\ra_\lam =\gamma_{n\lam}'
\label{eq:ensplit}
\eea
%
%\vspace{0.2cm}
% %%%%%%%%%%%%%%%%%%%%% figure %%%%%%%%%%%%%%%%%%%%%%%%%%%%
\begin{figure}
%\vspace{1cm}
\includegraphics[width=0.95\columnwidth]{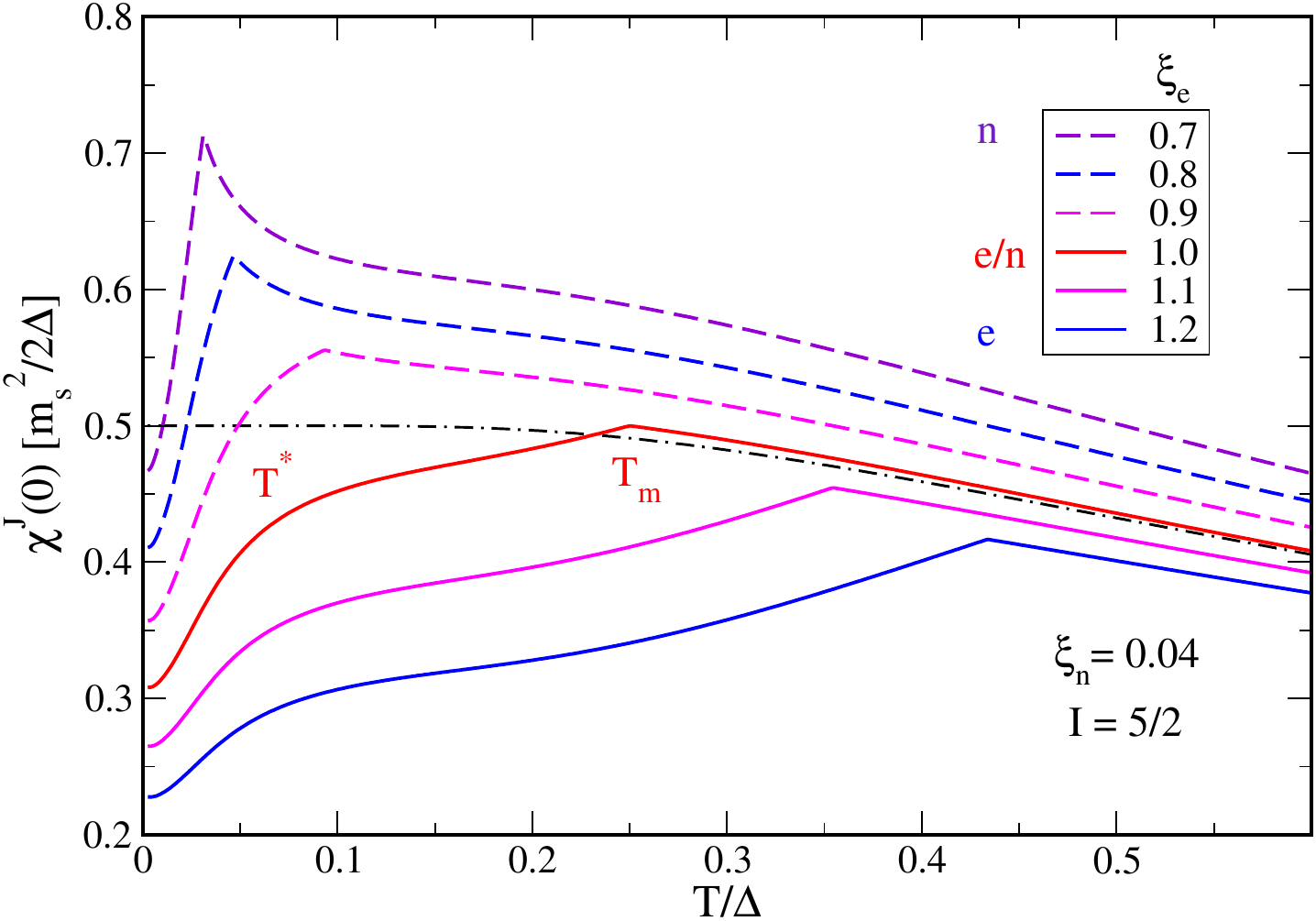}
\caption{{\BLU {\it Homogeneous} electronic  RPA susceptibility $\chi^J$(\bq=0) (Eq.~(\ref{eq:RPAhomsusz})) for AFM state for  $\xi_e$ below $(< 1)$ and above $(>1)$ critical value in the n- and e- dominated regimes, respectively. For the latter $T_m$ (cusp) and  $T^*\simeq \xi_n\De$ (hump) are separate energy scales that merge below the critical value of $\xi_e$. The cusp position for $\xi_e$ is tracing the $T_m(\xi_e)$ curve in Fig.~\ref{fig:fig2}. Black dashed line is reference of paramagnetic ($\xi_e=1$, $\xi_n=0$) state.}}
\label{fig:fig4}
\end{figure}
%%%%%%%%%%%%%%%%%%%%%%fig%%%%%%%%%%%%%%%%%%%%%%%%%%%%%%%
%
respectively, which are again normalised  according to $\hDe_{e\lam}=\De_{e\lam}/\De$ and $\hDe_{n\lam}=\De_{n\lam}/\De$. The former is always positive while the latter, which is caused by the hyperfine coupling to the induced moment,  may have both signs depending on the sublattice. This allows us also to present the $2\times (2I+1)$ split energy levels at each sublattice $\lam=A,B$ in the compact form
\bea
E_{\lam s}^\tau=\frac{\De}{2}(s\hDe_{e\lam}+2\tau\hDe_{n\lam})
\label{eq:eigen}
\eea
where the electronic two-singlet and the nuclear indices are $s=\pm,\tau=-I..I$, respectively.
%
%%%%%%%%%%%%%%%%%%%%%%%%%%%%%%%%%%%%%%%%
\begin{center}
\begin{table}
\caption{Magnetic characteristic quantities of 4f shell for non-Kramers
rare earth elements with J denoting ground state total angular momentum, $g_J$ the associated
Land\'e factor and $\mu_{eff}=g_J\sqrt{J(J+1)}$ the effective moment.
These stable rare earth isotopes  have 100\% abundance ratio  with resulting unique nuclear ground state spin I and  nuclear moment $\mu$ (with $\mu_n=e\hbar/2M_pc$ defining the nuclear magneton; $M_p$=proton mass) and hyperfine coupling constant A.}
\vspace{0.5cm}
\begin{center}
\begin{tabular}{c @{\hspace{4mm}} c @{\hspace{4mm}} c  @{\hspace{4mm}} c}
\hline\hline
RE$^{3+}$ ion \;\;\; & $^{141}$Pr \;\;\;\; & $^{159}$Tb \;\;\;\; & $^{169}$Tm \;\;\;\; \\ \hline
J & 4 &6 & 6 \\
g$_J$ &  $\frac{4}{5}$ & $\frac{3}{2}$  & $\frac{7}{6}$ \\
$\mu_{eff}$ [$\mu_B$] & 3.58 & 9.72& $7.56$  \\
\hline
I & $\frac{5}{2}$& $\frac{3}{2}$ & $\frac{1}{2}$  \\
$\mu [\mu_n]$   & 3.92 &  $1.52$ & $-0.20$ \\
A [mK]  & 52.5\cite{jensen:91} &  71.9\cite{kofu:13} & -18\cite{giglberger:67} \\
\hline\hline
\end{tabular}
\end{center}
\label{tbl:REdata}
\end{table}
\end{center}
%%%%%%%%%%%%%%%%%%%%%%%%%%%%%%%%%%%%%%%%%%
%
The coupled MF Eqs.~(\ref{eq:MFA}) for $\la S_x\ra$ and $\la I_z\ra$ have to be solved numerically for each nuclear spin size I as a function of control parameters $\xi_e$ and $\xi_n$ characterising the intersite electronic exchange and the on-site hyperfine coupling.

Examples for the temperature and field evolution of electronic and nuclear order parameters are shown in Figs~\ref{fig:fig3},\ref{fig:fig10}. For the FM the phase transition turns into a continuous crossover at $T_m$ while in the AFM it persists up to a critical field. The critical field curve $h_{cr}(T)$ is implicitly defined by the vanishing of the staggered moment $M_s=\fs(\la S_x\ra_A-\la S_x\ra_B)$ via the equation $M_s(h_{cr},T)=0$ and likewise this determines the field dependence of $T_m(h)$. 
The phase boundary is strongly influenced by the action of the hyperfine coupling as shown in Fig.~\ref{fig:fig9} and its resulting anomalous characteristics will be discussed further in Sec.~\ref{sec:discussion}.\\

{\BLU
At this point one should reflect to which degree the MFA-RPA treatment of the model gives reasonable and reliable
results, assuming we are dealing with a 3D lattice structure (which, as mentioned before, does not appear explicitly).
Corrections beyond the MFA would involve the quantum and thermal fluctuations governed by the excitation
spectrum of the model. According to the structure of the Hamiltonian in Eq.~(\ref{eq:Ham}) only the electronic pseudospins
have dynamics which lead to magnetic exciton bands that may be considered as dispersive singlet-singlet CEF excitations. Their dispersion, temperature and field dependence for the various singlet ground state models were investigated in detail
in \cite{thalmeier:21,akbari:23,thalmeier:24,thalmeier:25}. It was found that under most circumstances for AFM case with  $\bQ=(\pi,\pi,\pi)$ at $T_m(h')$ the spectrum has a finite gap given for both $\lam=A,B$ by \cite{thalmeier:24}:
\bea
\hspace{-0.2cm}\omega(\bQ,T_m(h'))=
\omega_0\bigl(\hDe^2_{eT_m}-1\bigr)^\fs(1-\xi_e^{-2}\hDe_{eT_m}^2)^\fs
\label{eq:softmode}
\eea
with $\omega_0=\De[(\xi_e\De)/(2T_m)]^\fs$ which is nonvanishing except for zero field where $\hDe_e(T_m)=0$ and at the critical field when $T_m(h'_c)=0$ and then  $\hDe_e(T_m)=\xi_e$. Then the phase boundaries in Fig.~\ref{fig:fig9} might only be affected at the endpoints by fluctuation contributions beyond MFA.
One has to keep in mind that on these two points the soft mode occurs only at the ordering wave vector \bQ~ and then for 3D lattice the momentum region for fluctuation contributions is small and in particular will not lead to diverging corrections. However, at zero field  the soft mode occurring at $T_m$ may modify the specific heat temperature dependence around the transition temperature. Furthermore when the excited state is not a singlet but a triplet the excitations  always stay gapped. The same is true for the singlet-singlet model when the contribution of higher lying states are included in the ordering temperature \cite{thalmeier:24}. Therefore in any realistic situation the critical magnetic exciton 'soft-modes' at the ordering wave vector that give fluctuation corrections to order parameters, phase boundaries and specific heat will be arrested  at a finite frequency gap. In fact in  all of the singlet ground state induced magnets that have been investigated by inelastic neutron scattering a true soft mode  has not been found. In such case the fluctuations will not qualitatively invalidate the molecular field picture used here but only lead to quantitative  modifications.  An actual computation of these modifications would, however, require a different approach beyond the scope of this work by using the Bogoliubov technique to calculate the magnetic exciton dispersion and their interactions \cite{thalmeier:21} to be able to treat their  bosonic statistics at low temperatures properly.}
% %%%%%%%%%%%%%%%%%%%%% figure %%%%%%%%%%%%%%%%%%%%%%%%%%%%
\begin{figure}
%\vspace{1cm}
\includegraphics[width=0.90\columnwidth]{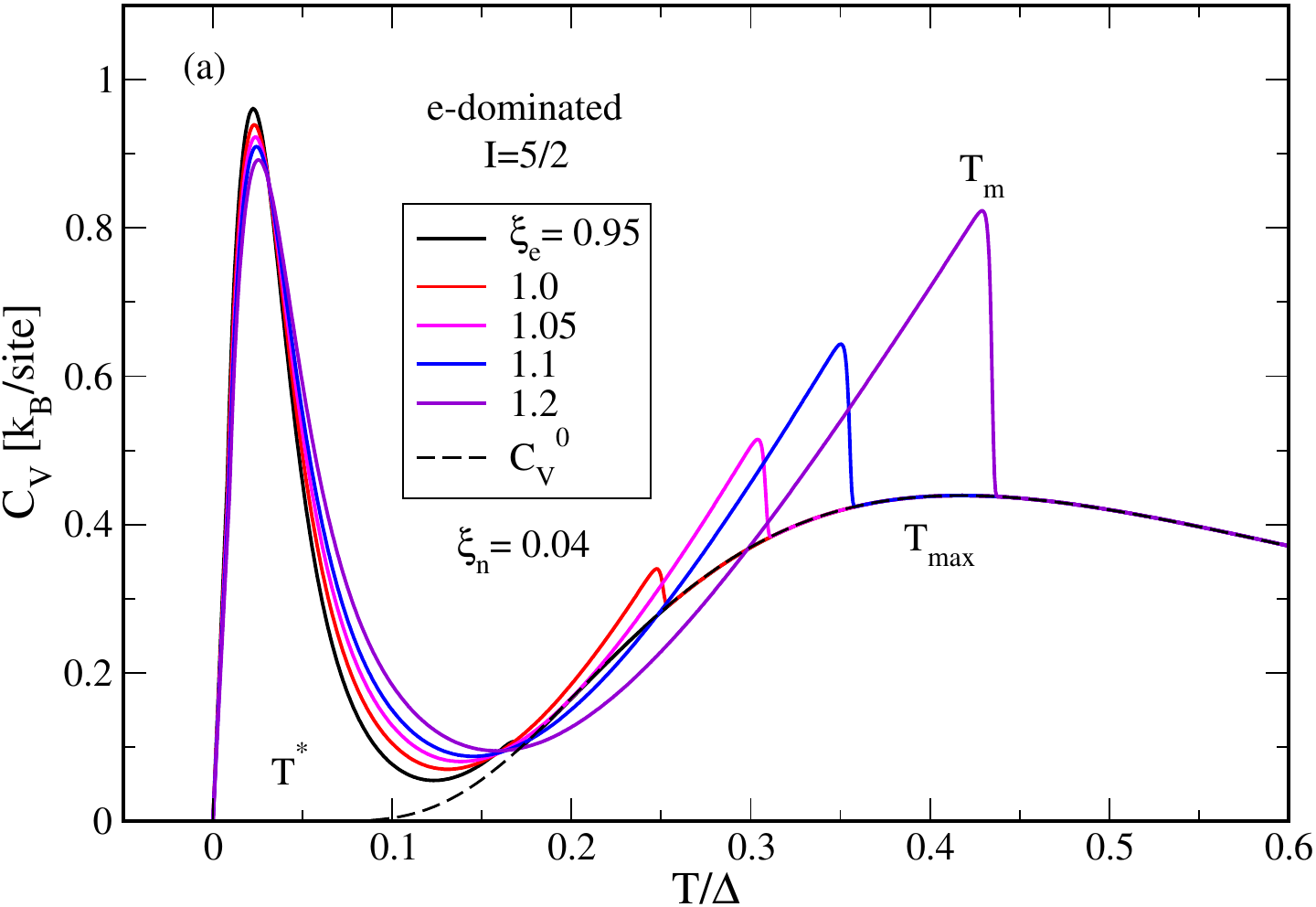}\\[0.2cm]
\includegraphics[width=0.90\columnwidth]{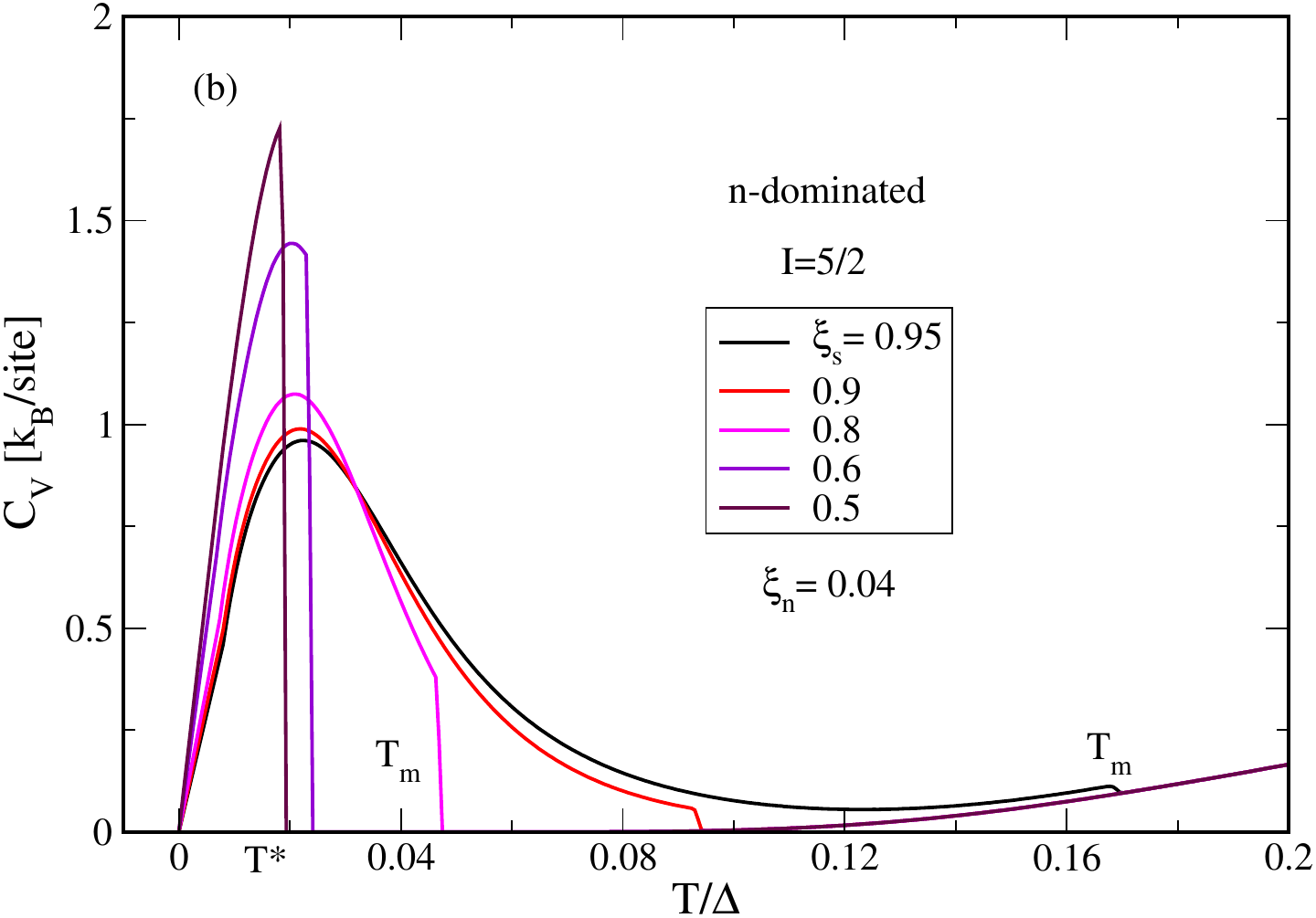}\\[0.2cm]
\includegraphics[width=0.90\columnwidth]{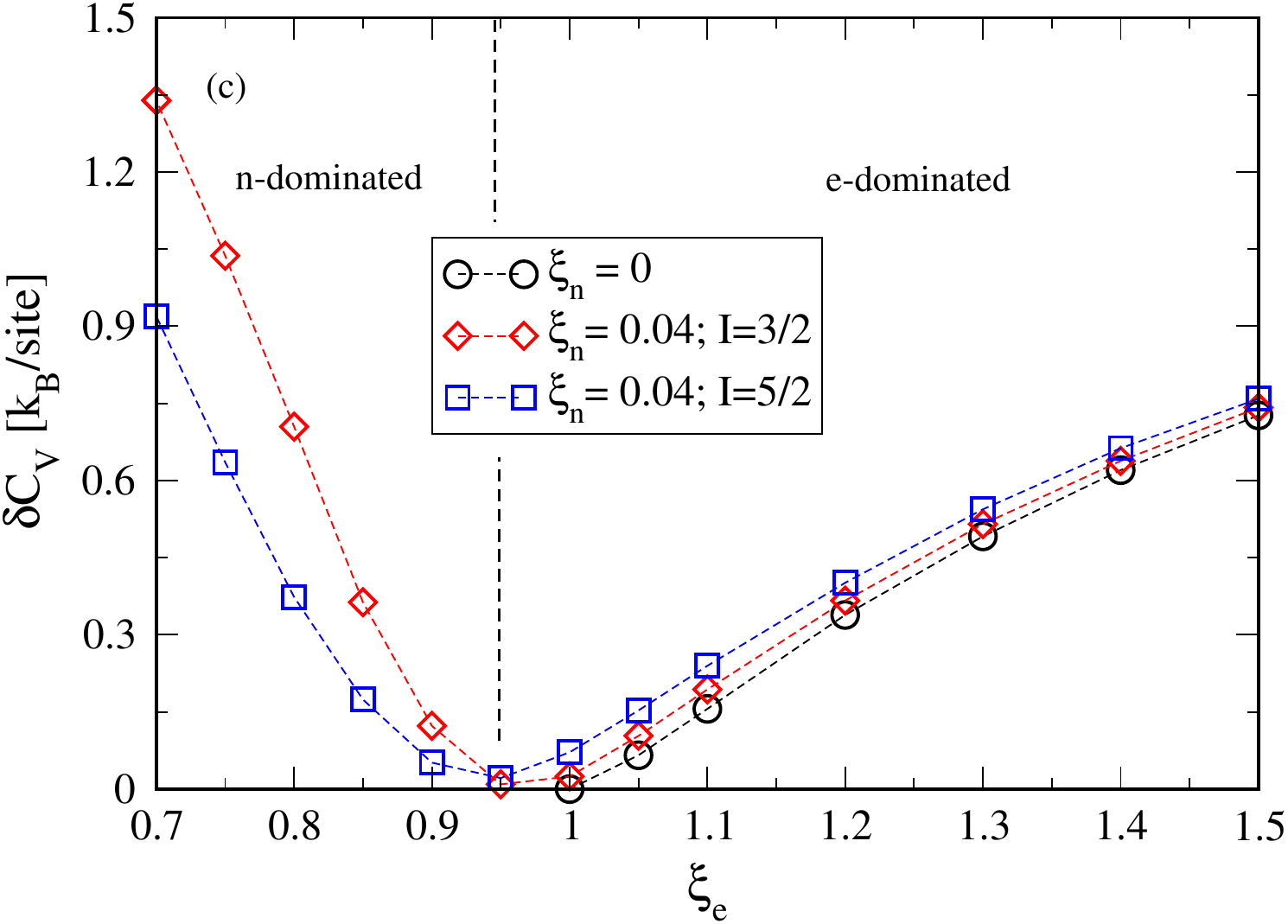}
\caption{{\BLU(a) Specific heat  
for various sub- and above critical electronic control parameters $\xi_e$ and  $I=\frac{3}{2}$ . A three-peak appearance
is displayed: i)  broad underlying CEF Schottky peak (dashed line) at $T_{max}$ due to the CEF splitting, ii) for $\xi_e >1$ a superposed induced ordering peak at T$_m$ and iii) a  low temperature nuclear specific heat peak with maximum at $T^*\simeq \xi_n\De$ corresponding to nuclear spin splitting appearing below $T_m$. (b) For subcritical $\xi_e$ but finite $\xi_n$ the ordering at $T_m$ shows crossover into the region dominated by the hyperfine coupling around $T^*$ (see Fig.~\ref{fig:fig2}).  (c)  Evolution of specific heat jump $\delta C_V(T_m)$ with nonmonotonic change from e-donminated $\xi_e >1$ (corresponding to (a))  to nuclear hyperfine n- dominated regime $\xi_e <1$ (associated with (b)).} }
\label{fig:fig5}
\end{figure}
%%%%%%%%%%%%%%%%%%%%%%fig%%%%%%%%%%%%%%%%%%%%%%%%%%%%%%%
%

\subsection{Susceptibility in the ordered regime}

The electronic and nuclear single site zero-field ($h$=$0$) susceptibilities in the ordered phase may also be derived using the transformed pseudo spin operator of Eq.~(\ref{eq:Strans}) and the expression in Eq.~(\ref{eq:statsus0}). We obtain for the electronic part $\chi_0^J(T)=\bigl(\frac{m_s^2}{2\De}\bigr)\hchi^J_0(T)$ with (cf. Eq.~(\ref{eq:ensplit}))
% $(\hDe_e=\De_e/\De)$
%
\bea
\hspace{-0.4cm}\hchi_0^J(T)=\frac{1}{\hDe_e^3}\tanh\frac{\De_e}{2T}
+(1-\frac{1}{\hDe_e^2})\bigl(\frac{\Delta}{2T}\bigr)\cosh^{-2}\frac{\De_e}{2T}
\label{eq:susJ}
\eea
which now contains a modified vanVleck and a pseudo-Curie term. Note that the latter
vanishes for small temperature despite the $(1/T)$ factor due to the other exponential population
factor. In the paramagnetic
phase $\De_e=\De$ and then the expression in Eq.~(\ref{eq:parsus0}) is recovered.
The nuclear spin susceptibility is now modified due to the splitting of nuclear spin states
and leads to another pseudo-Curie expression $\chi_0^I(T)=\bigl(\frac{2}{\De}\bigr)\hchi_0^I(T)$ with
\bea
\hchi_0^I(T)=\bigl(\frac{\Delta}{2T}\bigr)\bigl[\hP^C_n(T)-\la I_z\ra^2\bigr]
\label{eq:susI}
\eea
where an additional nuclear level population function is defined by 
\bea
\hP^C_n(T)&=& 2Z_n^{-1}\sum_{\tau=\fs}^I\tau^2\cosh\Bigl[\bigl(\frac{\De}{2T}\bigr)2\tau\hDe_n\Bigr]
\eea
On approaching T=0 the two terms in Eq.~(\ref{eq:susI}) approach $I^2$  and compensate leading to exponential suppression of the  Curie prefactor (1/T). In the paramagnetic state with $\la I_z\ra =0$ and $\tiP^C_n=\frac{1}{3}I(I+1)$ this reduces to Eq.~(\ref{eq:parsus0}).\\

{\BLU Using the above results and $I_{\bq=\bQ}=-I_e$ for the AFM case the homogeneous electronic RPA susceptibility for paramagnetic and ordered regime may then be expressed according to Eq.~(\ref{eq:susRPA}) as
\bea
\chi^J(\bq=0)=\frac{\bigl(\frac{m_s^2}{2\De}\bigr)\hchi_0^J(T)}
{1+[\xi_e-(2\xi_n)^2\hchi^I_0(T)]\hchi_0^J(T) }
\label{eq:RPAhomsusz}
\eea
}
The physical electronic susceptibility then has to be multiplied by $(g_J\mu_B)^2$. A similar nuclear RPA susceptibility
is obtained by replacing the numerator in the above equation with $(2/\De)\hchi_0^I$. The latter has a physical prefactor
 $(g_n\mu_n)^2$. Since the nuclear magneton $\mu_n\ll\mu_B$ it gives negligible contribution to the susceptibility.
 The temperature dependence of the homogeneous susceptibility in Eq.~(\ref{eq:RPAhomsusz}) is shown in Fig.~\ref{fig:fig4} 
 and discussed in Sec.~\ref{sec:discussion}.
 %
% %%%%%%%%%%%%%%%%%%%%% figure %%%%%%%%%%%%%%%%%%%%%%%%%%%%
\begin{figure}
%\vspace{1cm}
\includegraphics[width=0.95\columnwidth]{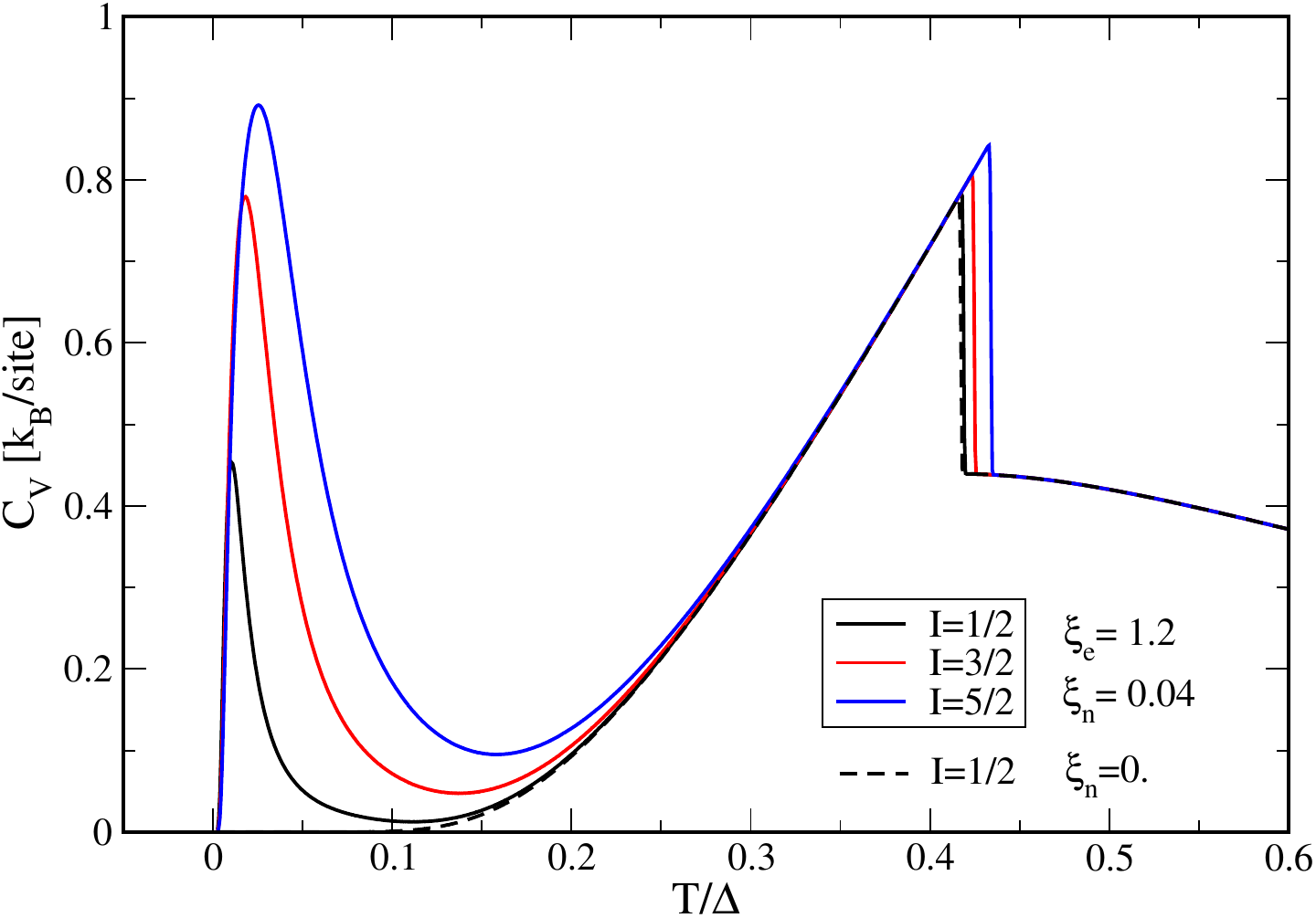}
\caption{ Specific heat curves for fixed control parameters $\xi_e,\xi_n$ and various nuclear spin size I which strongly influences the peak height due to entropy $\sim\ln(2I+1)$ contained in it. The dashed line corresponds to the CEF specific heat for $\xi_n=0$ (Eq.~(\ref{eq:sCV})).}
\label{fig:fig6}
\end{figure}
%%%%%%%%%%%%%%%%%%%%%%fig%%%%%%%%%%%%%%%%%%%%%%%%%%%%%%%
%
 %
% %%%%%%%%%%%%%%%%%%%%% figure %%%%%%%%%%%%%%%%%%%%%%%%%%%%
\begin{figure}
%\vspace{1cm}
\includegraphics[width=0.95\columnwidth]{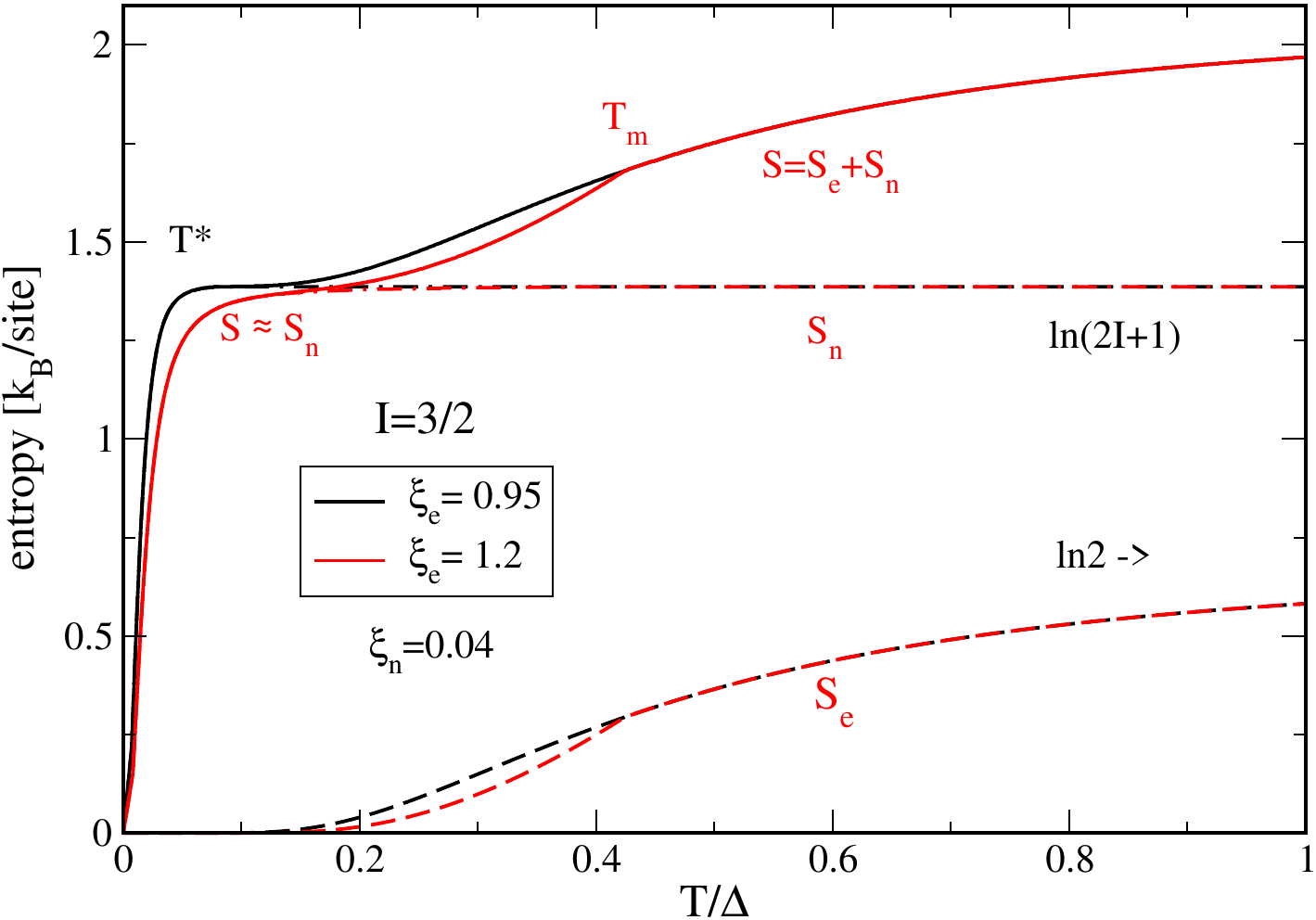}
\caption{Electronic $(S_e)$, nuclear $(S_n)$ and total entropy $S$ as function of temperature for two above (red)
and below (black)  critical control parameters. The entropy release for $S_e$ is mostly due to depopulation of singlets
but also due to induced ordering at $T_m$. The nuclear entropy is released due to the $(2I+1)$ nuclear level splitting caused by appearance of $\la S_x\ra$ at $T_m$. For subcritical $\xi_n=0.95$ the nuclear entropy is directly released by the nuclear dominated order around $T_m\approx T^*$ (cf. Figs.~\ref{fig:fig2},\ref{fig:fig5}(b)). }
\label{fig:fig7}
\end{figure}
%%%%%%%%%%%%%%%%%%%%%%fig%%%%%%%%%%%%%%%%%%%%%%%%%%%%%%%
%
\section{Electronic and nuclear parts of the specific heat, entropy and internal energy}
\label{sec:spec}
%
% %%%%%%%%%%%%%%%%%%%%% figure %%%%%%%%%%%%%%%%%%%%%%%%%%%%
\begin{figure*}
%\vspace{1cm}
\includegraphics[width=1.5\columnwidth]{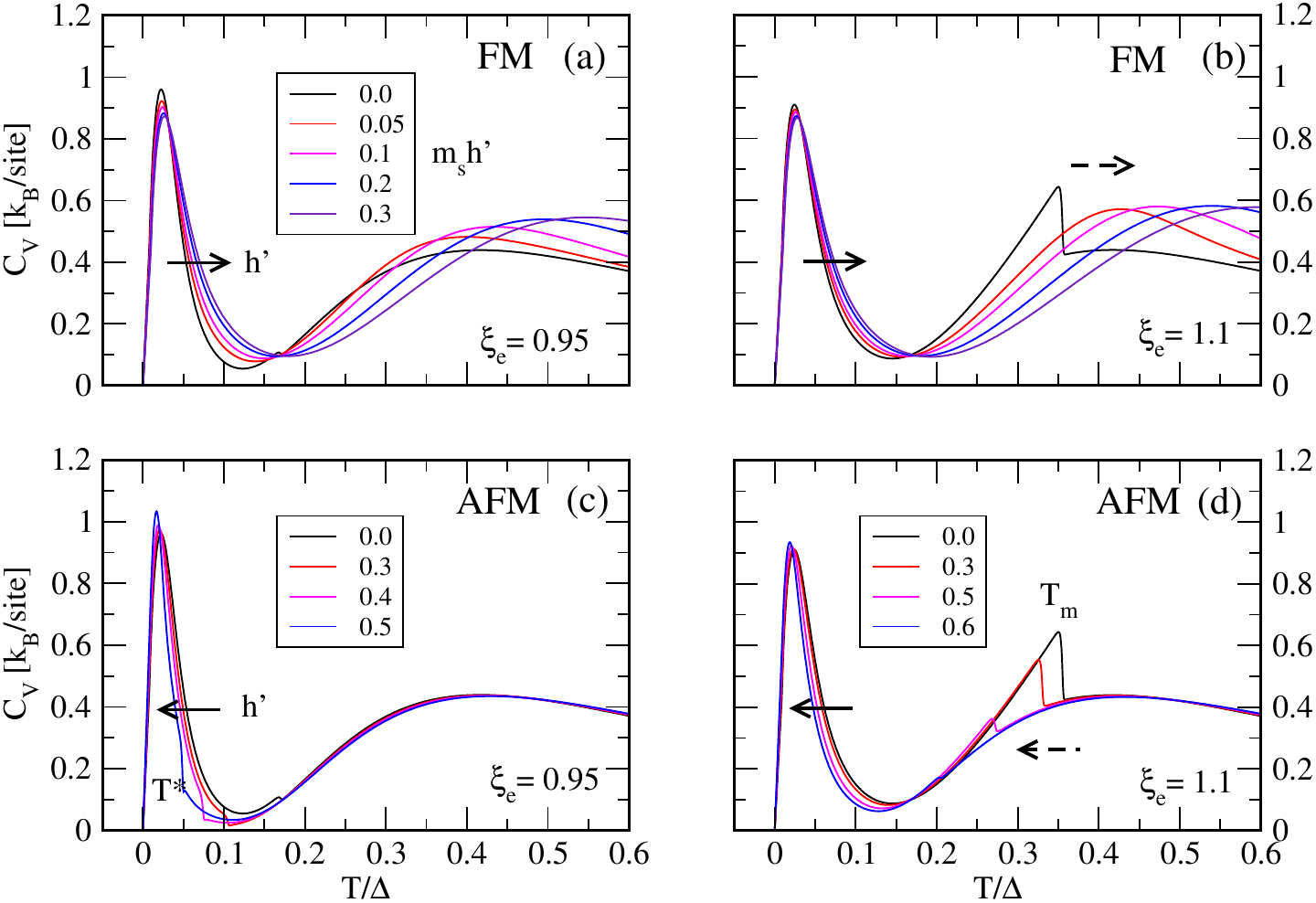}
\caption{Nuclear $(I=\frac{5}{2})$ and electronic specific heat evolution in external field for FM (a,b) and AFM(c,d) cases for sub- and above- critical control parameter $\xi_e$ and $\xi_n=0.04$ throughout. Dimensionless field strength $m_sh'$ is given in the boxes (equivalent for a,b). Evolution of anomalies with {\it increasing} $h'$ is indicated by arrows. In the FM case larger fields lead to a broadening of the nuclear peak (a,b) at $T^*/\De \simeq \xi_n$ (a,b)  and a rapid suppression, broadening and shift to higher temperature of the zero field jump at $T_m$ (b). In contrast for the AFM case the nuclear peak narrows with increasing field (c,d) while the AFM transition jump at $T_m$ (d) prevails but is reduced.}
\label{fig:fig8}
\end{figure*}
%%%%%%%%%%%%%%%%%%%%%%fig%%%%%%%%%%%%%%%%%%%%%%%%%%%%%%%
%

Of major interest is the signature of the electronic and nuclear contributions to the specific heat since
caloric experiments are most common in the low temperature regime in question. We approach this problem
by first calculating the thermodynamic potentials like entropy $S(T)$ and  internal energy $U(T)$. 
The entropy per sublattice site is given in terms of level occupation probabilities $p^\tau_{\lam s}(T)$ (Appendix \ref{sec:occfunc}) as 
\bea
S_\lam(T)=-\sum_{s\tau}p^\tau_{\lam s}\ln p^\tau_{\lam s}
\eea
which may be separated into an electronic and nuclear part $S_\lam=S_{e\lam}+S_{n\lam}$ (for noninteger I) according to
\bea
S_{e\lam}&=&-\frac{\De}{2T}\hDe_{e\lam} \hP_{e\lam}(T)+\ln\{2\cosh\Bigl[\bigl(\frac{\De}{2T}\bigr)\hDe_{e\lam}\Bigr]\}\\
S_{n\lam}&=&-\frac{\De}{2T}\hDe_{n\lam}\hP_{n\lam}(T)+
\ln\{ 2\sum_{\tau=\fs}^I\cosh\Bigl[\bigl(\frac{\De}{2T}\bigr)2\tau\hDe_{n\lam}\Bigr]\}\non
\label{eq:entropy}
\eea
Note that this expression is invariant against a constant shift of the combined level scheme.
The internal energy which is related by $U_\lam(T)=TS_\lam(T)-T\ln Z_\lam$ is obtained as
\bea
U_\lam(T)&=&-\frac{\De}{2}\hDe_{e\lam}\hP_{e\lam}(T)+\De\sigma\xi_e\la S_x\ra_\lam\la S_x\ra_{\blam}
%U_n(T)&=&U_n^0+U_n^{op}=-\frac{\De}{2}\hDe_n\hP_n(T)-\De(2\xi_n)\la S_x\ra\la I_z\ra\non
\label{eq:uint}
\eea
Formally the nuclear part has a similar structure. But in this case the first term
due to the nuclear spin splitting is exactly compensated by the second constant MF term describing 
the hyperfine interaction with the electronic moment. However, the nuclear degrees are implicitly incorporated in the electronic splitting $\hDe_{e\lam}(T)$ and $\la S_x\ra_\lam$. The total entropies and internal energies
(per lattice site) are then obtained by averaging over the sublattices A,B in the AFM ($\sigma=-1$) case, while the sublattice index $\lam$ may be ignored in the FM case $(\sigma =1)$ . Then the  specific heat is obtained from
the thermodynamical potentials via
\bea
C_V(T)=\bigl(\frac{\partial U}{\partial T}\bigr)_V =T \bigl(\frac{\partial S}{\partial T}\bigr)_V
\label{eq:specific}
\eea
by numerical differentiation which has been checked to lead to identical results for both expressions.
%The asymptotic form of the specific heat in the low temperature regime $T\ll T_m\ll\De$ is given in Appendix \ref{sec:asymptotic}.
In the case of zero hyperfine coupling $\xi_n=0$ only the electronic singlet degrees contribute and then the specific heat may be derived in closed form from the above equations as \cite{thalmeier:24}.
%
%\be
%\bl
\bea
\left\{
\begin{array}{l l}
C_V(T)=\bigl(\frac{\De_T}{2T}\bigr)^2\bigl(\cosh^2\frac{\De_T}{2T}-\frac{\De_0}{2T}\bigr)^{-1}\;\;\;& T\leq T_m\\[0.4cm]
C_V^0(T)=\bigl(\frac{\De}{2T}\bigr)^2\cosh^{-2}\frac{\De}{2T}\;\;\;&  T>T_m
\end{array}
\right.
\label{eq:sCV}
\eea
%\el
%\ee
%

for magnetic and paramagnetic phases, respectively, where we used the abbreviations $\Delta_T=\De\hDe_e(T)$ and $\De_0=\xi_e\De$.
$C_V^0(T)$ is the paramagnetic  background Schottky anomaly which peaks around $T_{max}=0.42\De$ for the two-singlet system.  It is shown as a reference by the dashed line in Fig.~\ref{fig:fig5}(a). $C_V(T)$  exhibits the jump at the induced moment transition temperature $T^0_m$. Expressions of the jump size at $T^0_m$ as function of control parameter $\xi_e$ are given in Ref.~\onlinecite{thalmeier:24}. It tends to zero when approaching the QCP from above, i.e. $\xi_e\rightarrow 1+\delta;\;\;\delta\ll 1$. The field and temperature dependence of the specific heat in the above general  expressions for finite $\xi_n$ in Eqs.~(\ref{eq:uint},\ref{eq:specific}) is contained in the excitation energies $\De^\lam_{e,n}$ and the order parameters $\la S_x\ra_\lam$, $\la I_z\ra_\lam$ obtained from the selfconsistent solution of MF equations Eq~(\ref{eq:MFA}).
The (magneto-) caloric properties of the model are presented in Figs.~\ref{fig:fig5},\ref{fig:fig6} and \ref{fig:fig8} and are discussed
in Sec.~\ref{sec:discussion}.

\section{Influence of nuclear electric quadrupole splitting}
\label{sec:quadru}

In the previous analysis we have assumed that the nuclear spin states are $(2I+1)$- fold degenerate.
However for $I\geq \frac{3}{2}$ an electric field gradient (EFG) at the nuclear site originating from the occupied
electronic states and resulting in a quadrupolar potential  will lift the degeneracy into pairs of time reversed 
states with $I_z=\pm\tau$. It is then to be expected to influence significantly the magnetic transition
temperature, in particular around and below the QCP at $\xi_e=1$ where the hyperfine interaction plays an important role.
The change in transition temperature due to the nuclear quadrupole splitting will now be investigated.\\
The quadrupolar Hamiltonian for nuclear spins reads \cite{abragam:61}
\bea
H^n_Q=\frac{\zeta}{4I(I-1)}\sum_i[3I_z^2(i)-I(I+1)]
\label{eq:HQ}
\eea
with the quadrupolar coupling constant $\zeta=\frac{e^2qQ}{h}$ where $Q$ is the nuclear quadrupole moment
and $eq=V_{zz}$ the uniaxial EFG. This causes a splitting of nuclear spin states already in the paramagnetic phase
which is added to the splitting caused by the molecular field in the induced moment phase. Together this leads to
multiplet energy levels given by an extension of Eq.~(\ref{eq:levelenergy0}) as
\bea
E_{\lam s}^\tau=\frac{\De}{2}[s\hDe_{e\lam}+2\tau\hDe_{n\lam}+\tau^2\hDe_Q]
\label{eq:levelenergy2}
\eea
where the last term describes the paramagnetic quadrupolar splitting with
$\hDe_Q=\De_Q/\De=3\zeta/[2I(I-1)\De]$ which can have both signs depending on the signs of the EFG, the nuclear quadrupole moment and on I. A constant shift which has no influence
on the thermodynamics has been omitted. The splitting will modify the nuclear spin susceptibility.
For this purpose we need to obtain the population factors of the split nuclear levels.
They are now given by
\bea
p_\lam^\tau&=&Z_n^{-1}\exp[-\frac{\De}{2T}(2\tau\lam\hDe_n+\tau^2\hDe_Q)]\non\\
Z_n&=&2\sum_{\tau=1/2}^I\exp(-\frac{\De}{2T}\tau^2\hDe_Q)\cosh\frac{\De}{2T}2\tau\hDe_n
\label{eq:occquadru}
\eea
The electronic level occupations and partition functions and therefore the susceptibility
$\chi_0^J(T)$  are unchanged because the
eigenstates are not modified by the Hamiltonian of Eq.~(\ref{eq:HQ}). With the above results
we can then obtain the quadrupolar modified nuclear spin susceptibility  as
\bea
\chi_0^I&=&\frac{\hP_n^Q(T)}{T}\non\\
\hP_n^Q(T)&=&2Z_n^{-1}\sum_{\tau=1/2}^I\tau^2\exp(-\frac{\De}{2T}\tau^2\hDe_Q)
\label{eq:quadsus} 
\eea
It is still of the Curie type due to splitting into nuclear spin doublets, however, the numerator
is no longer a Curie constant as in Eq.~(\ref{eq:parsus0}) but a temperature dependent occupation
function. It is interesting to consider three limiting cases  valid for any  $I\geq\frac{3}{2}$: i) For $T\gg |\De_Q|$ we have
$\hP_n^Q(T)\rightarrow \frac{1}{3}I(I+1)$ ii) for  $T\ll |\De_Q|$ and $\De_Q>0$ we obtain with $I_{eff}:=\fs$
$\hP_n^Q(T)\rightarrow \frac{1}{3}I_{eff}(I_{eff}+1)=\frac{1}{4} < \frac{1}{3}I(I+1)$.  iii) finally for  $T\ll |\De_Q|$ and $\De_Q<0$
we get $\hP_n^Q(T)\rightarrow I^2 >  \frac{1}{3}I(I+1)$
This means that depending on the sign of the quadrupolar splitting and its size as compared to 
the transition temperature the effective Curie numerator may be smaller or larger than the numerator $\frac{1}{3}I(I+1)$
for the degenerate nuclear spin states. Therefore a finite $\De_Q$ may lead to a decrease or increase in the nuclear ordering temperature for positive or negative sign, respectively. \\
%
% %%%%%%%%%%%%%%%%%%%%% figure %%%%%%%%%%%%%%%%%%%%%%%%%%%%
\begin{figure}
%\vspace{1cm}
\includegraphics[width=0.99\columnwidth]{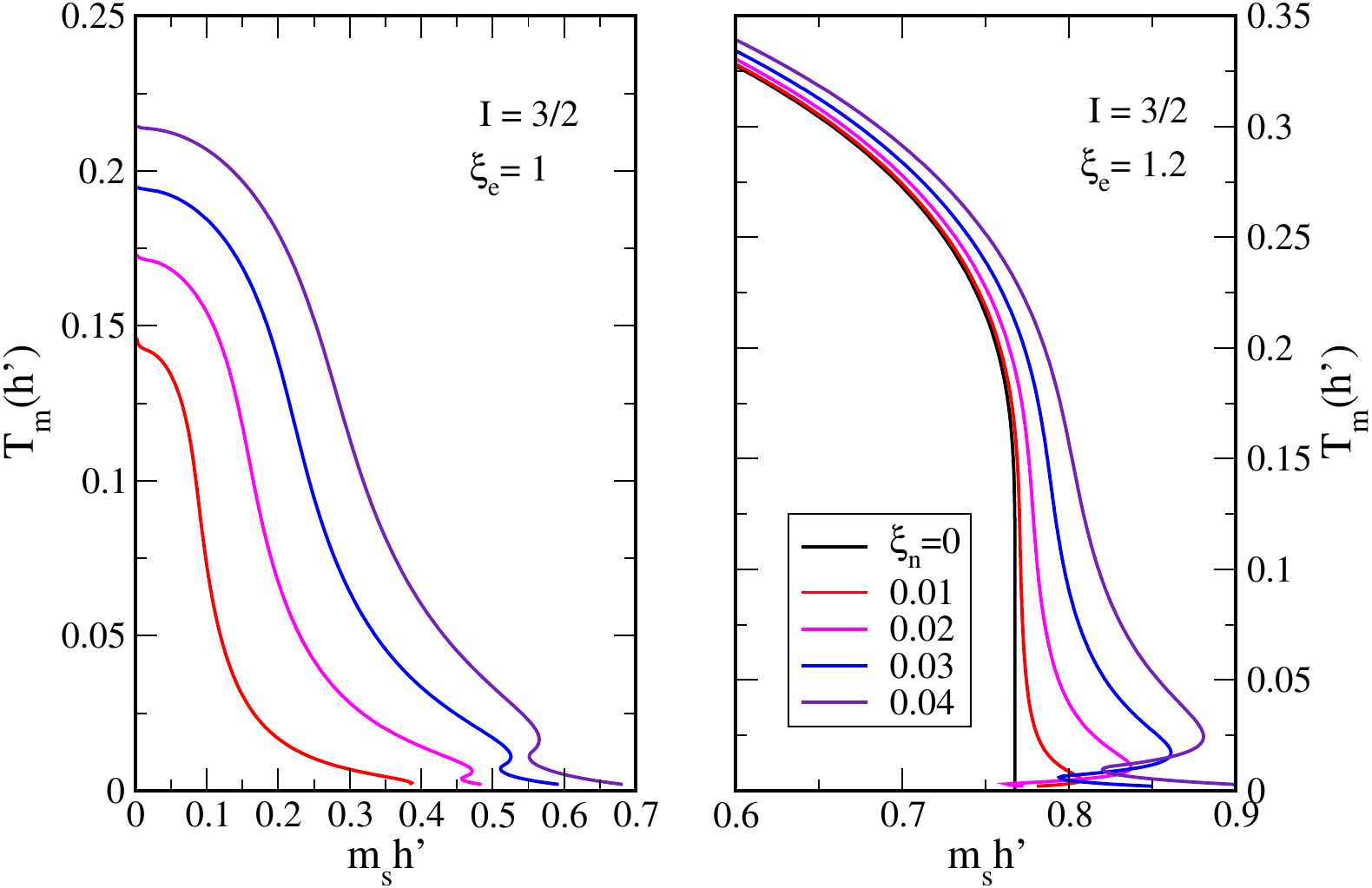}\\[0.2cm]
\includegraphics[width=0.90\columnwidth]{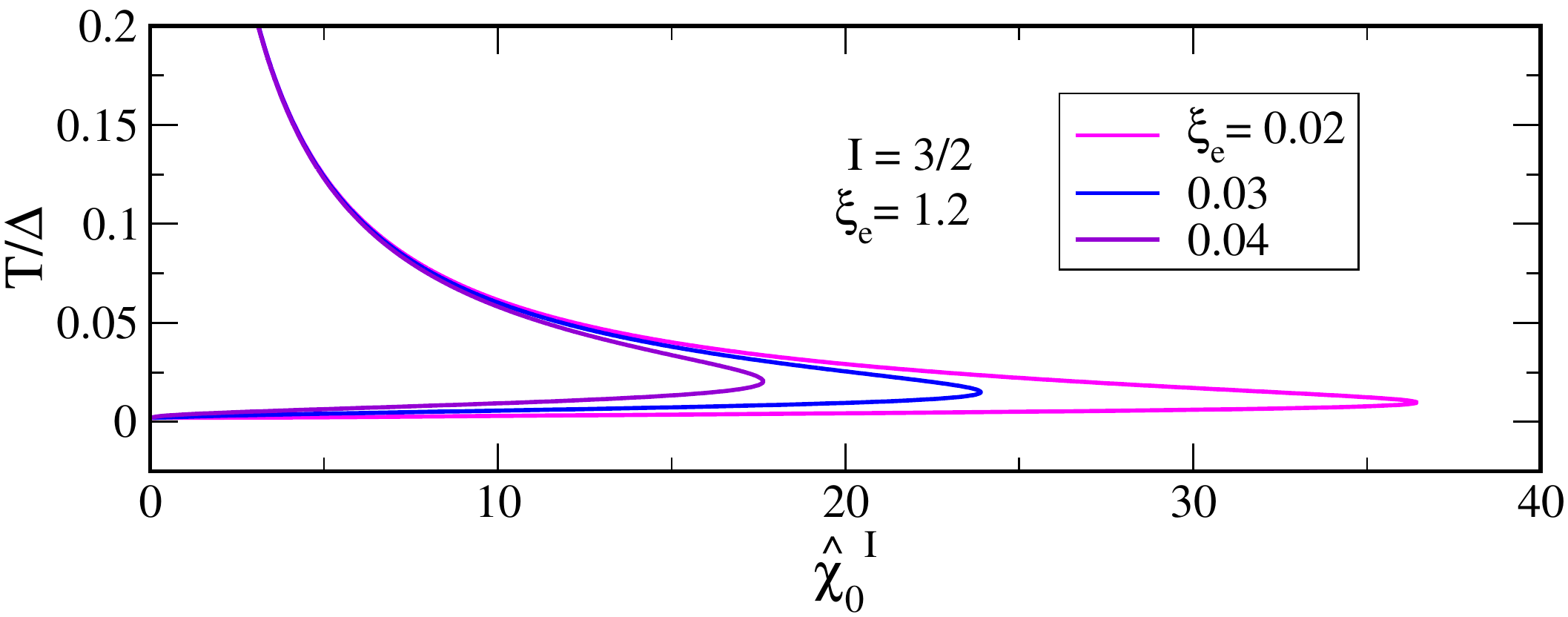}
\caption{Field dependence of the AFM transition temperature in the QCP regime (left) and the exchange dominated regime 
(right) {\BLU for various $\xi_n$. In the left panel $T_m$ increases rapidly with $\xi_n$ and develops reentrant behavior for large critical fields (low $T_m(h')$).  This becomes more pronounced in the right panel $(\xi_e >1)$, cf. Figs.~\ref{fig:fig10},\ref{fig:fig11}.  Bottom panel shows concomitant T- dependence of single-site nuclear susceptibility (Eq.~(\ref{eq:susI})) for field $m_sh'=0.95$ slightly above the critical value.}}
\label{fig:fig9}
\end{figure}
%%%%%%%%%%%%%%%%%%%%%%fig%%%%%%%%%%%%%%%%%%%%%%%%%%%%%%%
Inserting $\chi_0^I$ from Eq.~(\ref{eq:quadsus}) and the previous, unchanged  $\chi_0^J$
from Eq.~(\ref{eq:parsus0}) into the paramagnetic RPA susceptibility of Eq.~(\ref{eq:susRPA}) the condition for the divergence
of the latter leads now to a generalized implicit equation for the transition temperature $T_m(\xi_e,\xi_n,\hDe_Q,I)$
that contains the effect of the quadrupolar splitting. It is given by
\bea
\bigl[\xi_e+\hP_n^Q(T_m)(2\xi_n)^2\frac{\De}{2T_m}\bigl]\tanh(\frac{\De}{2T_m})=1
\label{eq:TmQ}
\eea
which may be obtained from the previous degenerate nuclear case given in Eq.~(\ref{eq:Tm})
simply by replacing the Curie constant $\frac{1}{3}I(I+1)$ by the effective 
temperature dependent Curie numerator in Eq.~(\ref{eq:quadsus}). From this it is clear that the quadrupolar
interaction dependence of $T_m(\hDe_Q)$ is limited by the curves for actual nuclear spin size $I$ and
quadrupolar ground state doublet with $\tau=\pm\fs$ or  $\tau=\pm I$ . This is shown in Fig.~\ref{fig:fig12} 
and discussed
in Sec.~\ref{sec:discussion}.\\

\section{Discussion of numerical results}
\label{sec:discussion}

The main focus of the previous analysis is firstly the dependence of the transition temperature or phase diagram of induced moment magnetism on the electronic and nuclear control parameters, in particular the removal of the induced moment QCP by the hyperfine coupling. Secondly the interplay and competition of various mechanisms on the entropy release witnessed by the specific heat anomalies, namely, CEF level depopulation, exchange dominated induced moment order and hyperfine dominated nuclear ordering. And thirdly the hyperfine coupling leads to anomalous nonmonotonic critical field dependence on temperature or equivalently a reentrance signature of magnetic order for large fields. The numerical results of these important effects of coupled CEF and nuclear degrees of freedom will now be discussed in detail for typical cases.\\
%
% %%%%%%%%%%%%%%%%%%%%% figure %%%%%%%%%%%%%%%%%%%%%%%%%%%%
\begin{figure}
%\vspace{1cm}
\includegraphics[width=0.99\columnwidth]{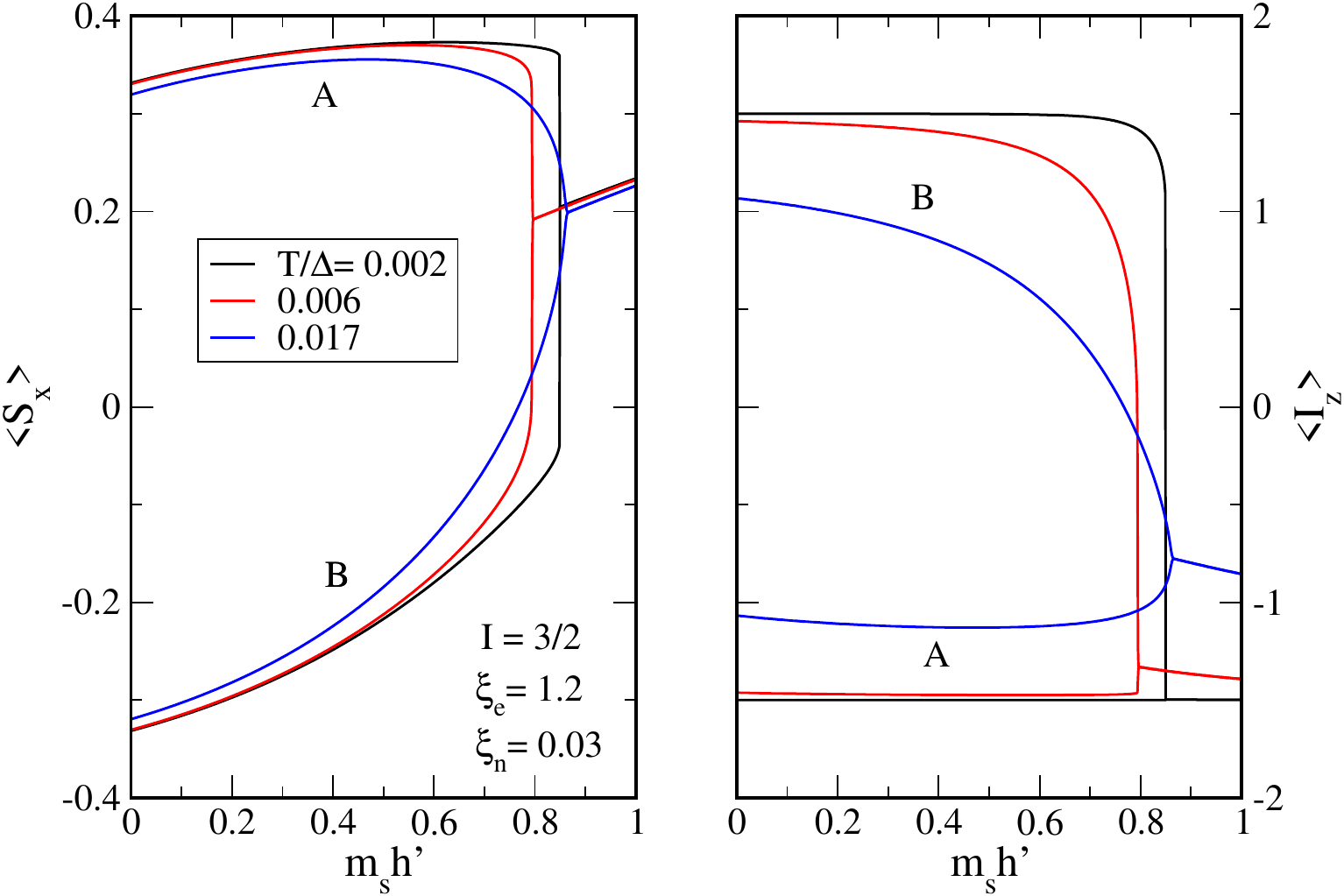}
\caption{Field dependence of electronic $\la S_x\ra$ and nuclear $\la I_z\ra$ order parameters  on A,B sublattices (left and right panels) in the low-T reentrant regime of the right panel in Fig.~\ref{fig:fig10}. {\BLU At the lowest T the transition appears first order like while it changes to  second order type when moving from  the sharp minimum of the critical field  to its maximum in Fig.~\ref{fig:fig9}. The critical field $h'_c(T)$ varies nonmonotonic with T.}}
\label{fig:fig10}
\end{figure}
%%%%%%%%%%%%%%%%%%%%%%fig%%%%%%%%%%%%%%%%%%%%%%%%%%%%%%%
%

{\BLU First we have to remark on the relevant size of dimensionless control parameters $\xi_e,\xi_n$ involved. These are combinations of microscopic energies and matrix elements (Eq.~(\ref{eq:control})) and therefore are not known apriori. In the case of e-dominated induced order with $\xi_e$ considerably above the QCP in Fig.~\ref{fig:fig2} the
transition temperature $T_m$ will be virtually independent of $\xi_n$  and may be obtained from inverting Eq.~(\ref{eq:Tm0})  to $\xi_e=1/tanh(\frac{\De}{2T_m})$ with $T_m$ and $\De$ the experimental transition temperature and singlet splitting,
respectively. The latter may possibly be obtained from inelastic neutron scattering or a global fit of the CEF level scheme to the high-temperature susceptibility. When $\xi_e$ is around the QCP or smaller the nuclear control parameter becomes essential
for $T_m$ and, knowing the experimental value of the latter Eq.~(\ref{eq:Tm}) provides only a constraint for the two parameters
which is visualized in  the contour plot $T_m/\De=const$ of Fig.~\ref{fig:fig2} (right panel)  where the possible values of $(\xi_e,\xi_n)$ form a 1D manifold. Our results suggest that in this (sub-)critical region the relative size of $\xi_e$ and $\xi_n$ could then be further constrained by analysis of the specific heat peak structure in Figs.~\ref{fig:fig5}(a-c) and the low temperature susceptibility (Fig.\ref{fig:fig4}) discussed below. Alternatively knowing A and $\De$ from experimental results and a calculated $m_s$ from a CEF model for the singlet states one can obtain $\xi_n$ and via Eq.(\ref{eq:Tm}) and/or Fig.~\ref{fig:fig2} one can deduce $\xi_e$ in the subcritical regime. An example for a Pr material is discussed in Appendix \ref{sec:PrCu2}.
}

For the numerical model calculations the electronic control parameter $\xi_e$ will be considered in the region of true induced moment order above the QCP $\xi_e=1$ but not too large and in particular for subcritical region $\xi_e < 1$ where the hyperfine interaction plays an important role. The nuclear control parameter $\xi_n=m_sA/2\De$ sets the hyperfine coupling in relation to the two-singlet CEF splitting. {\BLU The hyperfine coupling A itself can be determined by nuclear resonance spectroscopy \cite{abragam:61}. As mentioned, however, from the experimental $T_m/\De$ alone the nuclear control parameter $\xi_n$ can only be constrained in relation to $\xi_e$, but an estimate of its typical size can be given here.}
We take the CEF model in Eq.(\ref{eq:SSM}) with equal admixture coefficients corresponding to $\al=\pi/2$ which leads to the total angular momentum matrix element $m_s=8$. We assume a rather small singlet splitting of $\Delta=5$K in order to present the electronic and nuclear specific heat anomalies in graphs with linear rather than a distorted logarithmic temperature scale. This is not a restrictive assumption from the theory
but just convenient for the presentation. {\BLU In fact there are examples of singlet ground state systems with even smaller CEF splitting (Sec.~\ref{sec:materials})}. Using the typical size of $A\approx 50$ mK given in Table~\ref{tbl:REdata} for the potential singlet ground state candidate ions the nuclear control parameter is of the order $\xi_n\approx 0.04\ll 1$. The sign of A or $\xi_n$ has no influence because the spectra in Eqs.~(\ref{eq:eigen},\ref{eq:levelenergy2}) are invariant under $\xi_n,\tau \rightarrow -\xi_n, -\tau$. We therefore choose the positive sign. For the nuclear spin size we will only consider the noninteger cases $\text{I}=\fs,\frac{3}{2},\frac{5}{2}$ relevant from Table~\ref{tbl:REdata}. The possible nuclear quadrupolar splitting energy $\Delta_Q$ depends much on the nuclear environment {\BLU (via the size of the EFG) and the f-electron element (via the nuclear quadrupole moment) and is again obtained from nuclear spin spectroscopy. While it is commonly smaller or comparable to the hyperfine coupling $A$ for 4f nuclei, it may be considerably larger for the 5f case, i.e. for U-based singlet ground state magnets. Here we assume a range for the dimensionless $|\hDe_Q|=|\De_Q/\De|$ from zero to a maximum considerably  smaller than the hyperfine parameter $\xi_n$ and possibly of both signs.}\\

With the range of parameters constrained we first show the dramatic influence of hyperfine coupling on the ordering temperature which is presented by the phase boundary diagram in the $(\xi_e,\mbox{T})$ plane in Fig.~\ref{fig:fig2}. For vanishing hyperfine coupling there is a QCP for the intersite exchange induced moment order at $\xi_e=1$ and $T_m(\xi_e)$ collapses with a logarithmic singularity \cite{thalmeier:24} at this point leading to a paramagnetic state for $\xi_e<1$. Turning on a finite hyperfine coupling to the nuclear spins removes the QCP and the ordering temperature stays finite though small for all $\xi_e<1$ with a smooth crossover between the two regions. Naturally an increasing nuclear spin size leads to an increased $T_m$ for fixed $\xi_n$. This is also true when keeping spin size fixed and increasing $\xi_n$. For $\xi_e\rightarrow 0$ the ordering is dominated by the hyperfine coupling and the limiting value of $T_m$ is given below Eq.~(\ref{eq:Tm0}). Considerably above the QCP the hyperfine coupling has little influence on $T_m(\xi_e)$ which is intersite-exchange dominated in this regime. 
%
% %%%%%%%%%%%%%%%%%%%%% figure %%%%%%%%%%%%%%%%%%%%%%%%%%%%
\begin{figure}
%\vspace{1cm}
\includegraphics[width=1.05\columnwidth]{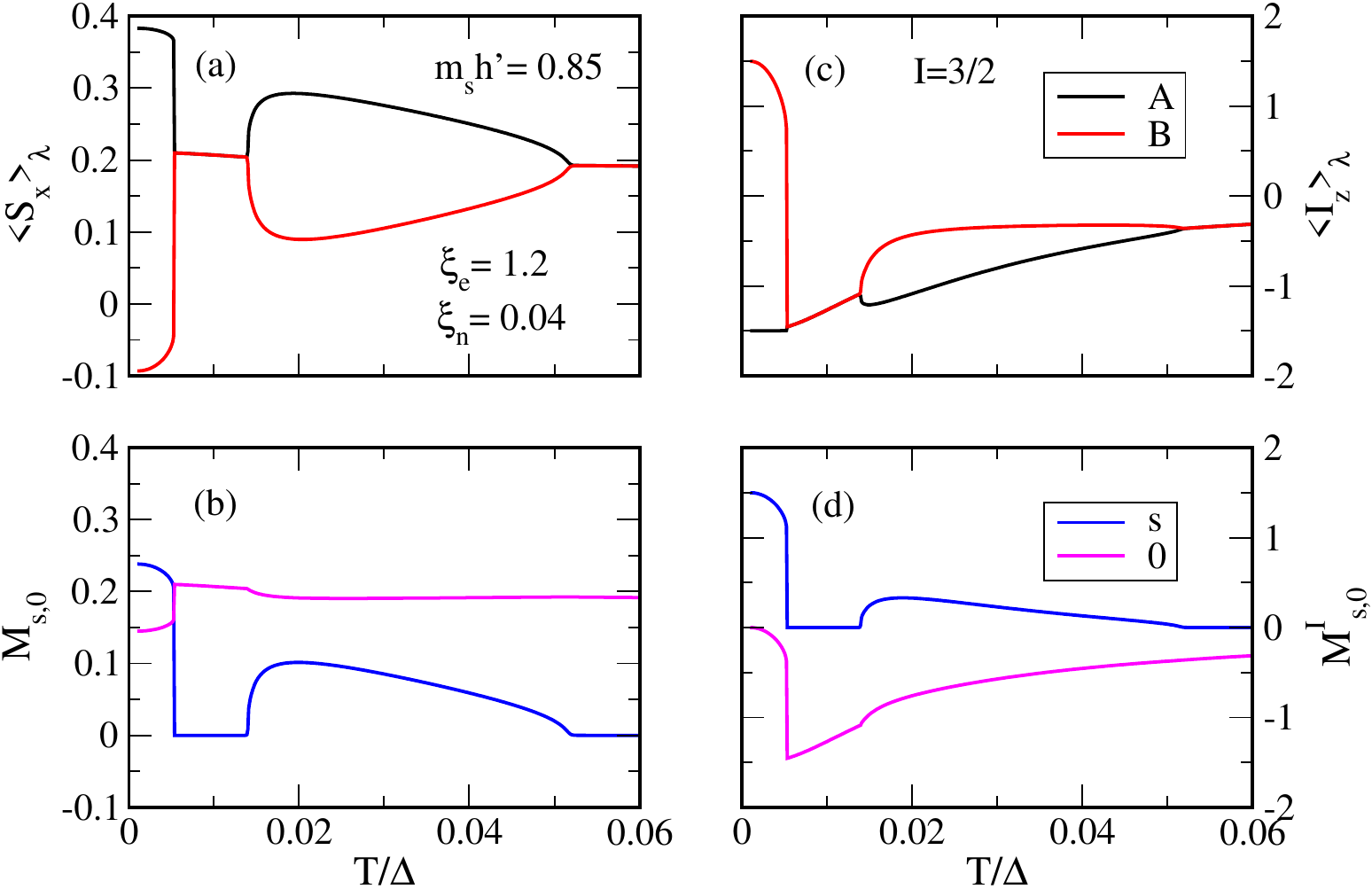}
\caption{{\BLU Complementary reentrant magnetic order as function of temperature for large  $m_sh'=0.85$ close to the maximum critical field in Fig.~\ref{fig:fig9}(b). Electronic (a) and nuclear (c) order parameters on sublattices A,B.
(b) Staggered AFM electronic order parameter is $M_s=\fs(\la S_x\ra_A-\la S_x\ra_B)$  and $M_0=\fs(\la S_x\ra_A+\la S_x\ra_B)$ 
is the homogeneous  moment.  Likewise for the nuclear case $M^I_{s,0}$ (with A$\leftrightarrow$ B). In an intermediate temperature regime the staggered order parameters reemerge.}}
\label{fig:fig11}
\end{figure}
%%%%%%%%%%%%%%%%%%%%%%fig%%%%%%%%%%%%%%%%%%%%%%%%%%%%%%%
%

The temperature dependence of the two (normalised) order parameters  below the transition temperature are shown in Fig.~\ref{fig:fig3}  for two typical control parameters above (a)  and below (b) the QCP. In the former case the transition at $T_m$ is CEF exchange dominated and $\la S_x\ra$ already saturates before the nuclear moment $\la I_z\ra$ becomes sizable only below
a temperature $T^*\approx \xi_n\De$ comparable to the nuclear spin splitting induced by the saturated $\la S_x\ra$ . Therefore the temperature dependence of both order parameters in Fig.~\ref{fig:fig3}(a) has a rather asymmetric appearance.
This is in contrast to  Fig.~\ref{fig:fig3}(b) for below-critical $\xi_e$ where the ordering at much lower temperature (Fig.~\ref{fig:fig2}) is dominated by the hyperfine coupling. In this case the nuclear and electronic moments evolve simultaneously below $T_m$.

The phase diagram in the $\xi_e$-T plane of Fig.~\ref{fig:fig2} has been obtained from the singularity of the paramagnetic RPA susceptibility in Eq.~(\ref{eq:susRPA}). It is also instructive to follow  the evolution of the latter  in the ordered regime. We consider the non-critical  {\it homogeneous}  susceptibility for the AFM state according to Eq.~(\ref{eq:RPAhomsusz}). {\BLU In the electronic induced moment regime for $\xi_e$ in the e-dominated case exhibits the typical  AFM cusp at $T_m$ and a lower separated hump and subsequent depression at $T^*\approx \xi_n\De$.  Below the critical value in the n-dominated regime the two energy scales merge to form a single cusp at $T_m$. The cusp as function of $\xi_e$ traces the $T_m(\xi_e)$ curve in Fig.~\ref{fig:fig2}. The size of cusp and hump and in particular their distance on the T axis
are a good indication whether one is in the e- or n- dominated regime of $(\xi_e,\xi_n)$ of the contours in Fig.2 (right panel)
for an experimentally known ratio $T_m/\De$.
For $\xi_e\geq1$ and absent hyperfine coupling $\xi_n$= 0 saturation \cite{thalmeier:24} at a finite low temperature value $[\xi_e(\xi_e^2+1)]^{-1}$ occurs, unlike the longitudinal susceptibility in a Kramers degenerate ground state (i.e. $\xi_e \gg 1$ ) magnet which tends to zero there. }

The zero-field specific heat of the model (Sec.~\ref{sec:spec}) is shown in Fig.~\ref{fig:fig5}(a) for $I=\frac{5}{2}$. It exhibits three superposed anomalies: i) the broad Schottky anomaly around $T_{max}$ due to the two-singlet splitting ii) the jump at the induced ordering  temperature $T_m$ for $\xi_e >1$ above the QCP value. iii) the nuclear specific heat peak at the temperature scale $T^*$. As the electronic control parameter is decreased through the QCP $T_m$ shifts to lower temperatures (cf. Fig.~\ref{fig:fig2}) and the specific heat jump is diminished. Finally for $\xi_e <1$ the ordering appears a much lower temperature $T_m\approx T^*$ dominated by the hyperfine coupling and the  jump at the ordering temperature appears inside the nuclear specific heat peak increasing again for lowering $\xi_e$ (in Fig.~\ref{fig:fig5}(b)). This intricate crossover appearance of the zero-field specific heat of Fig.~\ref{fig:fig5}(a,b) is also reflected in  Fig.~\ref{fig:fig5}(c) which demonstrates explicitly the 
nonmonotonic evolution of the specific heat jump across the QCP region.
Furthermore the dependence of specific heat anomalies on the nuclear spin I is shown in  Fig.~\ref{fig:fig6} for $\xi_e>1$ . The jump at $T_m$ is only weakly affected while the nuclear part at $T^*$ is strongly reduced by the smaller entropy release for decreasing I.

The associated entropy dependence on temperature is shown in Fig.~\ref{fig:fig7} on the two sides of the QCP. As mentioned before for $\xi_e>1$ part of the entropy release at high temperature  happens through excited singlet depopulation as well as through the induced order at $T_m$ and at low temperature the nuclear entropy is released via the nuclear spin splitting caused by the induced order parameter. For subcritical $\xi_e<1$ there is no more entropy release at $T_m$ and a clear plateau appears after the nuclear entropy has reached saturation $S_n=ln(2I+1)$ above the now hyperfine dominated ordering around $T_m\approx T^*$.
%
% %%%%%%%%%%%%%%%%%%%%% figure %%%%%%%%%%%%%%%%%%%%%%%%%%%%
\begin{figure}
%\vspace{1cm}
\includegraphics[width=0.99\columnwidth]{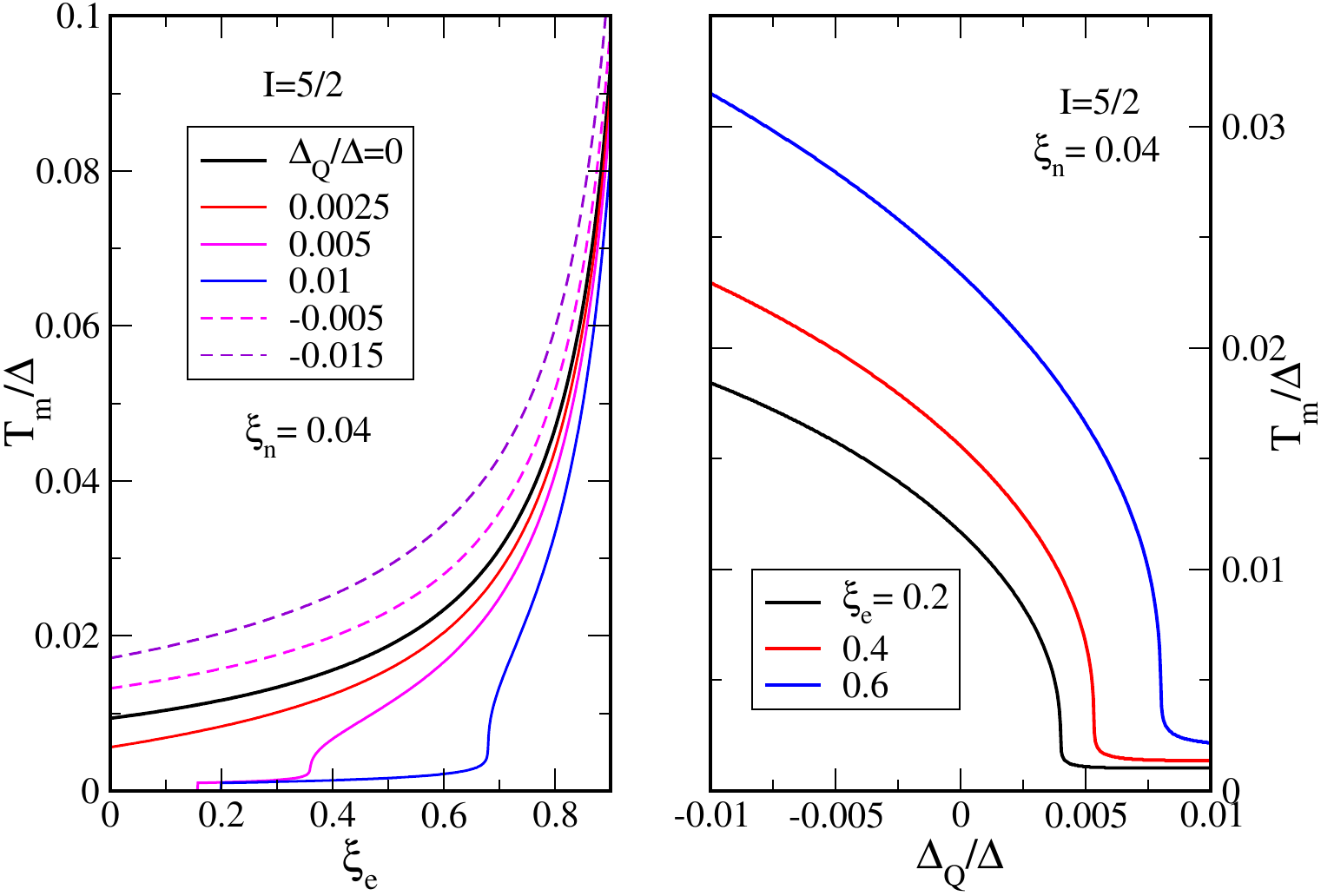}
\caption{{\BLU Left panel: Ordering temperature as function of $\xi_e$ for various positive and negative nuclear quadrupole coupling constants $\hDe_Q=\De_Q/\De$ in the nuclear dominated regime with subcritical $\xi_e < 1$.
Right panel: Complementary dependence of ordering temperature $T_m$ on the nuclear quadrupole coupling constant 
for subcritical $\xi_e$.}}
\label{fig:fig12}
\end{figure}
%%%%%%%%%%%%%%%%%%%%%%fig%%%%%%%%%%%%%%%%%%%%%%%%%%%%%%%
%
A further interesting aspect of the interplay of the three types of specific heat anomalies is their evolution with applied field.
In the model calculation only the Zeeman term for CEF singlets has been included due to $\mu_n\ll\mu_B$ but the nuclear spins  are indirectly influenced via the hyperfine coupling to the electronic order parameters $\la S_x\ra_\lam$ that depend on the field. The results for the field dependence for FM and AFM exchange coupling and below- and above- critical $\xi_e$ are shown in the four panels of Fig.~\ref{fig:fig8}. In the subcritical regime (a,c)  the broad Schottky anomaly shifts to higher temperature for FM while it is mostly unaffected for AFM. The nuclear peak around the  temperature $T^*$ broadens in the FM case while it narrows in the AF case. This behavior is caused by the quite different dependence of order parameters $\la S_x\ra_\lam$ and $\la I_z\ra$ on field strength and is also observed in the above-critical region of $\xi_e$ (b,d).
The higher temperature electronic part of the specific heat is much different in the FM and AFM cases \cite{thalmeier:25}. The jump at $T_m$ is quickly broadened and shifted to higher temperature for the FM case. On the other hand for AFM case the 
induced ordering and the associated specific heat jump prevails in finite field but is progressively suppressed in higher fields up to the critical field \cite{thalmeier:25}, thus approaching the subcritical appearance of the specific heat in (c).

The specific heat demonstrates that the FM transition changes to a gradual crossover  in a finite field because there is no more symmetry breaking involved. On the other hand the AFM transition persists up to a critical field. The corresponding H-T phase diagram as defined by the field dependence of the critical temperature is shown in the upper panels of Fig.~\ref{fig:fig9}. 
As mentioned before it is obtained by identifying the field strength where the staggered order parameter $M_s(T,h)$ vanishes.
Fig.~\ref{fig:fig9} demonstrates a reentrance behavior of the AFM magnetic transition temperature, most pronounced above
the critical $\xi_e$.  This corresponds to a nonmonotonic dependence of the critical field (the phase boundary) on temperature. It is caused by the rapid change of the electronic and nuclear order parameters and their coupling close to the critical field. This is demonstrated in the two panels of Fig.~\ref{fig:fig10}, both referring to the right panel of Fig.~\ref{fig:fig9}, which directly show the nonmonotonic shifting of the critical field with temperature. Equivalently the temperature dependence at fixed large field $h'=0.85$ in this region as shown in  Fig.~\ref{fig:fig11} clearly demonstrates the reentrance signature of magnetic order as witnessed in particular by the staggered order parameter $M_s(T)$ (red dashed) and its nuclear equivalent (right panel) in this figure. {\BLU For the model example given in the beginning of this section (and using $g_J=\frac{4}{5}$ for Pr$^{3+}$) we also give the absolute temperature and field scale for the critical field curves in Fig.~\ref{fig:fig9}. There $T_m/\De=0.35$ corresponds to T$_m$ = 1.75 K and $m_sh'=0.9$ to H = 1.05 T. These rather low scales are due to the low singlet splitting $\De=5$ K in this model.

 The reentrance behavior appears at the temperature scale  $T^*\simeq \xi_n\De$ corresponding to the nuclear spin splitting Eq.~(\ref{eq:ensplit}) around the critical field. In this region the homogeneous on-site nuclear susceptibility given by the pseudo-Curie expression in Eq.~(\ref{eq:susI}) exhibits a pronounced maximum (lower panel of Fig.~\ref{fig:fig9}) which destabilizes the staggered AFM order and leads to the minimum in the critical field curve around $T^*$ and the concomitant reentrance as function of temperature. This reentrance is stable and not subject to fine tuning since the peaked structure of the nuclear susceptibility  Eq.~(\ref{eq:susI})  caused by population/depopulation of nuclear levels is universal.} We note that this kind of reentrance  caused by hyperfine coupling is seen experimentally in a singlet ground state Pr-skutterudite compound \cite{bangma:23}. The details of the ordering are different for this compound because its low lying CEF states are of singlet-triplet type rather than two singlets and the ordered phase is field-induced quadrupolar type. In fact it has also been observed in Yb- compound with Kramers doublet ground state \cite{knapp:25} where it has been attributed to the influence of hyperfine coupling.\\

Finally we discuss the effect of possible quadrupole splitting of nuclear spin states  on the (zero-field) magnetic phase transition and how the phase boundary changes with the size of the quadrupolar nuclear coupling constant. {\BLU The latter we constrain to the regime $|\hDe_Q|\ll\xi_n$ appropriate for 4f elements. We expect it to be relevant only deep in the n-dominated regime with $\xi_e$ considerably below the QCP. This is shown in Fig.~\ref{fig:fig12} (left panel) with the critical temperature curves obtained from solving Eq.~(\ref{eq:TmQ}).} The full red curve gives the reference for zero quadrupolar potential $\hDe_Q$  under the presence of finite hyperfine coupling $\xi_n=0.04$. An increasing (positive) $\hDe_Q$ leads to a rapid characteristic suppression of subcritical $T_m(\xi_e)$, but never quite to zero. For above critical $\xi_e$ its influence is negligible. The overall shape of the critical temperature curve for the large $\hDe_Q$ (blue line) value approaches the behavior for zero hyperfine coupling. In contrast, for negative quadrupolar coupling constants $T_m$ increases and the shape of the (red) reference curve is preserved. This distinctive influence of nuclear quadrupole splitting is also seen in complementary right panel of Fig.~\ref{fig:fig12}  which plots directly the $\hDe_Q$ dependence for two subcritical $\xi_e$ and two nuclear spin sizes. The reason for the opposite dependence for different signs of $\hDe_Q$ was attributed in Sec.~\ref{sec:quadru}  to the different modification of the nuclear Curie-type susceptibility in the two cases.\\

{\BLU
\section{Remarks on candidate singlet ground state \lowercase{f}- electron materials}
\label{sec:materials}

This work is primarily intended as a model investigation to clarify the cooperation of electronic exchange and nuclear hyperfine coupling
in the induced moment mechanism without focusing on a specific compound. Nevertheless we give here a brief overview of f-electron compounds that qualify as induced moment systems  in the three possible regimes: above (e), around (e/n) and below (n) the QCP of Fig.~\ref{fig:fig2}(a).
The empirical criterion of this classification is given by the  experimentally determined ratio of $T_m/\De$  as obtained from inspection of this figure. We apply this to some known singlet ground state compounds (sometimes with degenerate first excited state) by giving the values of $T_m,\De$ and, wherever possible $\xi_e$ in Table~\ref{tbl:compound}. They can be roughly assigned to three groups:

{\BLN
(e) For $T_m/\De > 0.3$ one enters the e-dominated regime where the effect of hyperfine coupling is negligible.\\

(e/n) For $T_m/\De < 0.3$ but still sizable one is close to the QCP with $\xi_e\simeq 1^+$ (Fig.~\ref{fig:fig2}) and hyperfine coupling is becoming important in this regime.\\

(n) For $T_m/\De \ll 0.3$ one is in the n-dominated regime with $\xi_e <1$  and mainly hyperfine coupling responsible for induced ordering.\\
}

One has to keep in mind that the experimental determinations of the singlet ground state- excited state splittings have commonly a considerable margin, depending on the method used, therefore these classifications are not unambiguous. The above listings, however,  indicate that the nuclear dominated induced ordering is primarily observed among the Pr compounds as indeed has been realized already before \cite{andres:73}. Of these material examples PrCu$_2$ in the n-dominated regime is considered in more detail in Appendix \ref{sec:PrCu2}.}
%
%%%%%%%%%%%%%%%%%%%%%%%%%%%%%%%%%%%%%%%%
\begin{center}
\begin{table}
{\BLN
\caption{Collection of data for CEF singlet ground state compounds (the first excited CEF state may not be a singlet) with induced order where FM, AFM denotes the ferro-/antiferromagnetic case, IC  incommensurate magnetic order and HO hidden multipolar order. Here $T_m$ is the ordering temperature, $\Delta$ the splitting of lowest two CEF levels and $\xi_e$ the electronic control parameter.
The electronic dominated, intermediate and nuclear dominated regimes of induced order are denoted by e, e/n and n, respectively.}
\vspace{0.5cm}
\begin{center}
\begin{tabular}{c @{\hspace{4mm}} c @{\hspace{4mm}} c @{\hspace{4mm}} c @{\hspace{4mm}} c @{\hspace{4mm}}  c @{\hspace{4mm}} c}     
\hline\hline
regime & compound & state& T$_m$ & $\De$ [K] & $\xi_e$ & Ref. \\ \hline
e& PrNi & FM & $21$ K& 26 & $1.8$ & \cite{savchenkov:19}\\
& TbSb  & AFM  & $15$ K & $12$ & $ 2.7$ & \cite{cooper:70} \\
&UGa$_2$ & FM & $125$ K & $38$ & $6.6$ & \cite{marino:23}  \\
&URu$_2$Si$_2$ & HO & $17.5$ K & $35$ & $1.3$ & \cite{haule:10}  \\
\hline
e/n& Pr$_3$Tl& FM & $11.6$ K & $60$ &1.011 &\cite{birgeneau:72, andres:72,buyers:75} \\
& UPd$_2$Al$_3$& AFM & $14.3$ K & $66$ &1.015 &\cite{thalmeier:02} \\
& Tb$_3$Ga$_5$O$_{12}$ & AFM & $0.25$ K & $2.55$ & $\simeq 1$ &\cite{wawr:19,hammann:73} \\ \hline
n& PrCu$_2$& FM & $54$ mK & $2.9$ & $<1$ &\cite{kawarazaki:95} \\
& dhcp Pr& IC & $50$ mK & $41$ & $<1$ &\cite{moller:82} \\
& PrNi$_5$& FM & $0.4$ mK & $47$ & $<1$ &\cite{kubota:80,radwanski:00} \\
& PrCu$_6$& FM & $2.45$ mK & $20$ & $<1$ &\cite{miki:92,sugiyama:98} \\
\hline\hline
\end{tabular}
\end{center}
\label{tbl:compound}
}
\end{table}
\end{center}
%%%%%%%%%%%%%%%%%%%%%%%%%%%%%%%%%%%%%%%%%%
%

\section{Summary and conclusion}
\label{sec:summary}

In this work we investigated the influence of the hyperfine coupling with nuclear spins on the properties of CEF singlet-singlet induced moment magnets in the context of a highly symmetric Ising-type model that allows a mostly analytical treatment. Furthermore we include the effects of an external field and the influence of nuclear quadrupole potential leading to a splitting of nuclear spin states.\\

We have shown that the QCP existing in the purely electronic model is suppressed by the hyperfine coupling and replaced 
by a crossover from  electronic exchange CEF induced moment order with large critical temperature $T_m$ to a subcritical regime below the QCP where the order at a much lower $T_m$  is dominated by the hyperfine coupling and comes to lie
within the temperature regime of nuclear spin splitting energy.

Furthermore the temperature and field dependence of both electronic and nuclear order parameters in both ordered regimes was calculated showing characteristic differences in their appearance.
In particular the specific heat, its temperature, field and control parameter dependence was investigated. It was found
that it is determined by the interplay of three anomalies which is also reflected in the entropy release. They correspond to Schottky type anomaly due to upper singlet depopulation, specific heat jump at the induced moment ordering for above critical electronic control parameter and large and narrow nuclear peak due to lifting of nuclear spin degeneracy. The specific heat jump exhibits a pronounced nonmonotonic evolution when crossing the QCP region. The field dependence shows characteristic differences for these  anomalies for FM/AFM models and below/above critical control parameter. In particular the nuclear specific heat peak is broadening for the FM case and in contrast shows narrowing in the
AFM case, reflecting the distinct field dependence of the order parameters and associated excitation gaps in the two cases. Importantly the resulting H-T phase diagram was found to exhibit a peculiar reentrance signature connected with nonmonotonic critical field dependence on temperature which becomes more pronounced with increasing hyperfine coupling.\\
In addition we studied in detail the effect of a possible nuclear quadrupole coupling which lifts the nuclear spin degeneracy. It was found that a positive quadrupolar potential suppresses the transition temperature curve
while the negative quadrupole coupling enhances the ordering temperature preserving the shape of the phase boundary. This distinction may be traced back to the properties of the pseudo-Curie nuclear spin susceptibility in the two cases.

{\BLU Finally we may summarize the main achievements of this investigation:
We have given a treatment of electronic nuclear coupling and its influence on induced order 
for general nuclear spin size and a calculation of the systematic evolution of two- and three- peak appearance
of the total specific heat as function of electronic and nuclear control parameters
and nuclear spin size. Furthermore the systematic evolution of the electronic/nuclear specific heat peak structure in a magnetic field for FM as well AFM order was clarified and most importantly we identified  a reentrance behavior of the critical field curves and the associated behavior of electronic order parameter and nuclear spin ordering. It is the consequence of competition between hyperfine coupling to induced order and homogeneous nuclear polarization. In addition the qualitative influence of the quadrupolar nuclear potential  on the induced moment  transition temperature was demonstrated.}\\
The present work provides a solid theoretical framework for the discussion of the influence of nuclear spins with hyperfine coupling on singlet-singlet induced moment magnetism which may occur in compounds with non-Kramers f-electron ions with sufficiently low site symmetry of the crystalline electric field.\\[1.0cm]

%\newpage
{\BLU
\appendix
\section{Partition functions and thermal occupation functions}
\label{sec:occfunc}
For the calculation of order parameters, susceptibilities and thermodynamic potentials we
need the thermal occupations of the split electronic/nuclear level scheme under the presence of (sublattice) order parameters and an external field. In this appendix we define them for the case without
quadrupolar splitting. The latter is included in Sec.~\ref{sec:quadru}.\\

The thermal occupations of individual levels are given by
\bea
 p^\tau_{\lam s}=Z^{-1}_\lam\exp(-E^\tau_{\lam s}),
 \eea
 where $Z_\lam$ is the partition function for sublattice $\lam=A,B$ and $E^\tau_{\lam s}$ are the MF energy levels given by Eq.~(\ref{eq:eigen}). Because the Ising type hyperfine interaction
does not mix the products of electronic and nuclear states the partition function factorises according to
$Z_\lam=Z_{e\lam}\cdot Z_{n\lam}$, respectively.
\bea
\label{eq:partition}
Z_{e\lam}&=&2\cosh\Bigl[\bigl(\frac{\De}{2T}\bigr)\hDe_{e\lam}\Bigr]\\
Z_{n\lam}&=&
\left\{
\begin{array}{rl}
1+2\sum_{\tau=1}^I\cosh\Bigl[\bigl(\frac{\De}{2T}\bigr)2\tau\hDe_n\Bigr] \; &I\;\mbox{integer}\non\\
 2\sum_{\tau=\fs}^I\cosh\Bigl[\bigl(\frac{\De}{2T}\bigr)2\tau\hDe_{n\lam}\Bigr]
 \;& I\;\mbox{noninteger}
\end{array}
\right.
\eea
%
%%%%%%%%%%%%%%%%%%%%%%%%%%%%%%%%%%%%%%%%
The form of $Z_{n\lam}$ for {\it noninteger} nuclear spin $I=\fs,\frac{3}{2} ...$ is the relevant one 
for the isotopes of $4f$ elements which we have in mind here (Table~\ref{tbl:REdata}).
Then  the differences of thermal populations appearing in the MF equations (Eq.~(\ref{eq:MFA})) for the order parameters may be derived as
\bea
\label{eq:popdiff}
\hP_{e\lam}(T)&=&\tanh\frac{\De}{2T}\hDe_{e\lam}\non\\
\hP_{n\lam}(T)&=& 4Z_{n\lam}^{-1}\sum_{\tau=\fs}^I\tau\sinh\Bigl[\bigl(\frac{\De}{2T}\bigr)2\tau\hDe_{n\lam}\Bigr]
\label{eq:popfunc}
\eea
%
%Here the sum starts at $\tau=1$ or $\fs$ for I being integer or noninteger, respectively, with the corresponding
%nuclear partition function taken from Eq.~(\ref{eq:partition}).
These population functions depend implicitly on the induced moments $\la S_x\ra_\lam$ and $\la I_z\ra_\lam$ for each sublattice via the excitation energies $\hDe_{e\lam}, \hDe_{n\lam}$ which have a sublattice $\lam$- dependent size for finite field $h$. In zero field the sublattice dependence reduces to just a sign difference 
given by  $P_{n\lam}(T)= sign(\lam)P_n(T)$ with $sign(\lam)=\pm 1$ for $\lam=A,B$.

\section{Special case with $I=\fs$}
\label{sec:specialcase}

For the special case with nuclear spin $I=\fs$ the selfconsistency equations for the electronic and nuclear order parameters in Eq.~(\ref{eq:MFA}) assume a more transparent and explicit form which is given by
\bea
\hspace{-0.2cm}\la S_x\ra_\lam&=&\fs\frac{1}{\hDe_{e\lam}}\bigl\{m_sh'-2\xi_e\la S_x\ra_{\blam}\non\\
&&+\xi_n\tanh[(\frac{\De}{2T})(2\xi_n\la S_x\ra_\lam)]\bigr\}
\tanh[(\frac{\De}{2T})\hDe_{e\lam}]\non\\
\hspace{-0.5cm}\hDe_{e\lam}&=&\Bigl\{1+\Bigl(m_sh'-2\xi_e\la S_x\ra_{\blam}\non\\
&&+\xi_n\tanh[(\frac{\De}{2T})(2\xi_n\la S_x\ra_\lam)]\Bigr)^2\Bigr\}^\fs\non\\
\la I_z\ra_\lam&=&-\fs\tanh[(\frac{\De}{2T})(2\xi_n\la S_x\ra_\lam)]
\eea

\section{Control parameters for P\lowercase{r}C\lowercase{u}$_2$ singlet state magnet}
\label{sec:PrCu2}

In the beginning of Sec.~\ref{sec:discussion} we introduced the parameters for a hypothetical tetragonal $(J=4)$ CEF
singlet-singlet system defined in Eq.~(\ref{eq:SSM}). It was chosen to have the maximum magnetic matrix element $m_s=8$ between the two singlets. Then, for the small CEF splitting $\Delta$ assumed the nuclear control parameter
using the hyperfine coupling A of Pr (Table \ref{tbl:REdata}) is  obtained as $\xi_n=0.04$ . 

For real compounds the singlet-singlet model is mostly realised for low symmetry like orthorhombic $D_{2h}$ (where only singlet CEF states can occur). Then the difficulty of estimating $\xi_n$ stems from the fact that $m_s$ is hard to know since it depends critically  on numerous $D_{2h}$ CEF potential parameters that determine the composition of the two singlet wave functions. Nevertheless
we attempt this for the material example of PrCu$_2$ which belongs to the n-dominated class described in Sec.~\ref{sec:materials}  with very low T$_m=54$ mK (AFM). Its CEF potential and singlet state wave functions
were proposed in \cite{ahmet:96}. For the two lowest singlets they are approximately given by combinations of $|M\ra=|JM\ra$ free ion states as
\bea
|0\ra&=& a|0\ra  + b(|+4\ra+|-4\ra)\non\\
|1\ra &=& c(|1\ra+|-1\ra)+d(|3\ra+|-3\ra)
\eea
with the coefficients $a=0.8433, b=3797, c=0.3296, d=-0.6720$. (Here we applied a gauge change by a factor (-1) to
the $|1\ra$ singlet). Then the matrix element for the \bJ~operators
may be computed and it is largest for x-direction (easy axis of AFM order). Defining now the pseudospin via $J_x=m_sS_x$ it is given by $m_s=4(\sqrt{5}ac+\sqrt{2}bd)$ from which we obtain $m_s=1.14$. Using the experimental
singlet splitting \cite{kawarazaki:95} $\De=2.9 K$ and A of Table \ref{tbl:REdata} we get $\xi_n=m_sA/(2\De)=0.01$ which corresponds to the low value used in most previous figures. This is due to the relatively small value of $m_s$ for these two singlets. As explained in Sec.~\ref{sec:hyperfine} for CEF  uniaxial symmetry with easy z-axis matrix elements up to $m_s=8$ may be achieved which leads to correspondingly higher $\xi_n$ values provided the splitting of singlets stays sufficiently small. With the knowledge of  $\xi_n=0.01$ and using numerical solution of Eq.~(\ref{eq:Tm}) (Fig.~\ref{fig:fig2}) for the experimental ratio $T_m/\De=0.019$ of PrCu$_2$ we obtain the $\xi_e=0.97$ subcritical value for this compound.
}
%%%%%%%%%%%%%%%%%%%%%%%%      References        %%%%%%%%%%%%%%%%%%%%
%\newpage
%\bibliographystyle{prsty}
\bibliography{References}
\end{document}